\documentclass[journal=jctcce,manuscript=letter]{achemso}
\usepackage{geometry}

\SectionNumbersOn
\usepackage[T1]{fontenc} % Use modern font encodings

\usepackage{siunitx}
  \DeclareSIUnit{\calorie}{cal}
  \DeclareSIUnit\angstrom{\textup{\AA}}
  \DeclareSIUnit\hartree{\text{\ensuremath{E}}_{\mathrm{h}}}
  \DeclareSIUnit\walkers{\text{\ensuremath{N}}_{\mathrm{w}}}

\usepackage{xcolor}
  \definecolor{mygray}{rgb}{0.5,0.5,0.5}
\usepackage{colortbl}
\usepackage{booktabs}

\setkeys{acs}{
  maxauthors=10,
  articletitle=true,
  chaptertitle=true,
  doi=true,
}

\usepackage{enumerate}
\usepackage{amsthm}
  
\usepackage{mathtools}

\usepackage{blkarray}
\usepackage{cool}
  \Style{DSymb=\mathrm{d}, DShorten=true}
  \Style{IntegrateDifferentialDSymb=\mathrm{d}}
  \Style{ISymb=\mathrm{i},
         ESymb=\mathrm{e}}
\usepackage[version=4]{mhchem}
\usepackage{algorithm}
\usepackage{algpseudocode}
\usepackage{braket}
\usepackage{csquotes}

\usepackage{tikz}
  \usetikzlibrary{arrows.meta}
  \usetikzlibrary{calc}
  \usetikzlibrary{quotes}
  \usetikzlibrary{shapes.arrows}
  \usetikzlibrary{shapes.geometric}
  \usetikzlibrary{decorations.markings}
  \usetikzlibrary{decorations.pathmorphing}
  \usetikzlibrary{positioning}

\pgfarrowsdeclaredouble{<<s}{>>s}{Straight Barb}{Straight Barb}%   double stealth
\tikzset{
    cgvertex/.style={
        semicircle,
        shape border uses incircle,
        inner sep=1.25pt,
        draw
    },
    vertex/.style={
        circle,
        inner sep=1.5pt,
        fill=black,
        semithick
    },
    singlearrow/.style={
        postaction=decorate,
        decoration={
            markings,
            mark=at position .55 with {\arrow{Straight Barb}}
        }
    },
    doublearrow/.style={
        postaction=decorate,
        decoration={
            markings,
            mark=at position .6 with {\arrow{>>s}}
        }
    },
    3jsymbol/.style n args={3}

}
\author{Arta A. Safari}
  \affiliation[MPI FKF]{Max-Planck-Institute for Solid State Research, 70569 Stuttgart, Germany}
  \email{a.safari@fkf.mpg.de}
\author{Nikolay A. Bogdanov}
  \affiliation[MPI FKF]{Max-Planck-Institute for Solid State Research, 70569 Stuttgart, Germany}
  \email{n.bogdanov@fkf.mpg.de}

  \title{Effective Hamiltonians from Spin-Adapted Configuration Interaction}

\begin{document}
\singlespacing
% \begin{tocentry}

% Some journals require a graphical entry for the Table of Contents.
% This should be laid out ``print ready'' so that the sizing of the
% Inside the \texttt{tocentry} environment, the font used is Helvetica
% 8\,pt, as required by \emph{Journal of the American Chemical
% The surrounding frame is 9\,cm by 3.5\,cm, which is the maximum
% permitted for  \emph{Journal of the American Chemical Society}
% graphical table of content entries. The box will not resize if the
% content is too big: instead it will overflow the edge of the box.
%
% This box and the associated title will always be printed on a
% separate page at the end of the document.
%
% \end{tocentry}

% NOTE: The abstract is commented out for a reason.
%       Please leave it like this!
\begin{abstract}
A generalised extraction procedure for magnetic interactions using effective
    Hamiltonians is presented that is applicable to systems with more than two sites
    featuring local spins $S_i \geq 1$.
To this end, closed, non-recursive expressions pertaining to chains of
    arbitrary equal spins are derived with the graphical method of angular
    momentum.
The method is illustrated by extracting magnetic couplings from
    ab initio calculations on a \ce{[CaMn3^{(IV)}O4]} cubane.
An extension to non-sequential coupling schemes proves conducive to
    expressing additional symmetries of certain spin Hamiltonians.

\end{abstract}
% \pagebreak

\section{Introduction}
Transition metal compounds are a common motif in nature, with examples ranging
from single-molecule magnetism\cite{caneschi1991,sessoli1993,chibotaru2023,lee2017,titis2023,bogani2008,rinehart2011} to
    enzymatic catalysis\cite{guo2024,pamies2003,koehler2012,wang2013} and
    superconductivity\cite{li2019,takahashi2008,bednorz1986,lee2006}.
Their thermally accessible states are typically characterised by localised spins
    that can be rationalised in terms of phenomenological spin models like the
    Heisenberg-Dirac-Van Vleck Hamiltonian.
If the model admits analytical solutions, energy differences can be fitted to
    data from experiment or calculation in order to obtain effective interaction
    constants.
While a correspondence of the respective ab initio states to the model can be
    inferred through evaluation of the orbital-resolved spin correlation
    function\cite{sharma2012,dobrautz2021} or point group
    symmetry\cite{bastardis2007}, an exact
    map may not be established, because of the intrinsic difference between
    ab initio and model spins.\cite{griffith1960,mcweeny1965,malrieu2013}
A more rigorous method is the effective Hamiltonian approach that renormalises
    the full Hamiltonian onto the magnetic subspace.\cite{maurice2009,malrieu2013}
In this procedure, extraneous wave function components are
    eliminated, which enables an exact map to the model.

To obtain the effective Hamiltonian, hereinafter also called numerical matrix,
    eigenvectors of the full Hamiltonian are projected onto the model
    space and must be expressed in a compatible basis.
Different choices are possible; for instance, a $S_i = \frac{1}{2}$ Heisenberg
    model in the uncoupled basis corresponds to a wave function in Slater
    determinants.
Blocks of interest may then be targeted by choosing a total
    spin projection $m_S$ common to all considered states.
Alternatively, if blocking by total spin $S_{\text{tot}}$ is desired, the
    transformation to the coupled basis is given by Clebsch-Gordan
    coefficients.

Theories formulated in a basis of total spin
    eigenfunctions,\cite{shavitt1981,duch1985,matsen1964,kaplan2013} are well-suited to the description
    of magnetic interactions, since variational degrees of freedom related to
    recoupling of angular momenta can be treated separately from ones due to mixing
    of spatial configurations.
The resulting compression over spin multiplets has proven particularly beneficial
    for sparse configuration interaction (CI) solvers like the Density Matrix
    Renormalisation Group (DMRG)\cite{keller2016,sharma2012,wouters2014} or
    Graphical Unitary Group Approach Full CI Quantum Monte Carlo
    (GUGA-FCIQMC)\cite{dobrautz2019}.
Already a number of successful applications to systems consisting of more than
    two sites with large magnetic moment has been
    reported\cite{han2023,kurashige2013b,sharma2014b,limanni2021b,dobrautz2021};
    however, to the best of our knowledge, the effective Hamiltonian approach has
    not yet been formulated in this context.
Instead, analysis proceeded via local spin expectation values or DMRG
    entanglement diagrams.\cite{kurashige2013b,sharma2014b,dobrautz2021}
The purpose of this letter is threefold: (i) to explain the problem
    faced in the construction of effective Hamiltonians when more than two sites
    with $S_i \geq 1$ are considered, (ii) to demonstrate how recoupling
    transformations can be used to resolve this issue, deriving closed expressions
    for an arbitrary number of equal spins and (iii) to highlight the increased
    sparsity offered by nonstandard representations of the symmetric or
    unitary group.

Alternatively, it is possible to work with fragmentation
    methods\cite{liu1980,parker2013,nishio2019,agarawal2024} that expand the full
    wave function in a product basis of states with well-defined subsystem quantum
    numbers.
The resulting solutions already contain components that correspond to the
    uncoupled basis of spin models, making the construction of effective
    Hamiltonians straightforward, as briefly discussed for the localised active
    space method\cite{hermes2019} in the Supporting Information.

\section{Conversion between Coupling Schemes}
When two arbitrary magnetic moments on
    sites $A$ and $B$ are coupled, the resulting states are uniquely
    characterised by their total spin $S_{\text{tot}}$, e.g.
    $\frac{3}{2} \otimes \frac{3}{2}$ yields $S_{AB} =
    \{0 \oplus 1 \oplus 2 \oplus 3\}$.
For a two site-model, the Heisenberg Hamiltonian is diagonal in the coupled
    basis, allowing the derivation of interaction constants from energy differences
    using Land\/{e}'s interval rule.
When another spin $S_C$ is added, the reduction of the tensor product space is
    generally no longer multiplicity free, i.e. multiple states can have the
    same $S_{\text{tot}}$.
For example, adding $S_C = \frac{3}{2}$ to the case above results in:
    \begin{equation}
        \{0 \oplus 1 \oplus 2 \oplus 3\} \otimes \frac{3}{2} =
            \left(2 \cdot \frac{1}{2}\right) \oplus \left(4 \cdot \frac{3}{2}\right)
            \oplus \left(3 \cdot \frac{5}{2}\right) \oplus \left(2 \cdot
            \frac{7}{2}\right) \oplus \frac{9}{2}.
        \label{eq:cg-trinuclear}
    \end{equation}
To resolve this multiplicity problem, states are labelled with
    two quantum numbers that specify their genealogy, here
    $\ket{S_{AB}, S_{tot}}$.
The model matrix is blocked by $S_{tot}$, but not guaranteed to be
    diagonal, unless $S_{AB}$ is a good quantum number.

\begin{figure}[H]
    \includegraphics[width=0.49\textwidth]{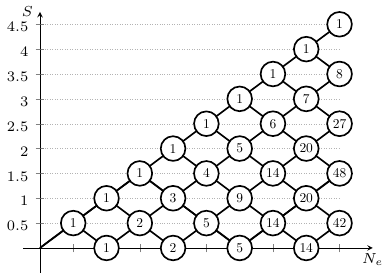}
    \includegraphics[width=0.49\textwidth]{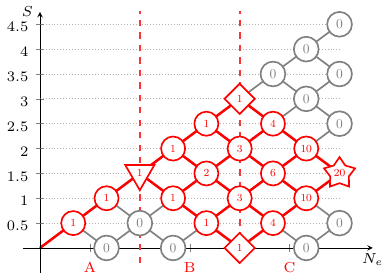}
    \caption{Left: Genealogical branching diagram in standard form for nine
    electrons. Right: Localising and site-ordering the orbitals expresses
    Hund's rule on the first site, marked by a triangle and
    reduces the number of possible paths.
    Shown in red are the remaining non-vanishing CSFs with $S_{tot} = \frac{3}{2}$.}
    \label{fig:gbd-compression}
\end{figure}

Whereas solutions to the Heisenberg model are characterised by both local spin
    quantum numbers $S_{A}$, $S_{B}$, $S_{C}$ and $S_{tot}$, it is more common in
    quantum chemistry to couple individual electrons, $s_i = \frac{1}{2}$, into a
    cumulative spin up to particle $j$, denoted $S_i^j$.
The resulting CSFs correspond to standard representations of the symmetric or
    unitary group:
\begin{equation}
    \ket{C} = \ket{S_1^1 S_1^2 \ldots S_1^n}.
\end{equation}
In the left side of Figure \ref{fig:gbd-compression}, the genealogical graph
    resembling this coupling scheme is shown for nine electrons in nine
    spatial orbitals.
It was shown that for magnetic system described by Heisenberg-like models,
    further sparsity may be introduced into the graph, by taking advantage of the
    stabilisation arising from parallel alignment of site spins due to Hund's
    rule.\cite{limanni2020a}
Localising and site-ordering the orbitals makes the first three electrons
    correspond to site $A$, marked by a triangle in Figure
    \ref{fig:gbd-compression}, and energetically
    penalized non-Hund CSFs with $S_A \neq \frac{3}{2}$ get effectively separated.
This sparsity manifests in the Full CI expansion as well by reducing the number of CSFs
    with appreciable weight.\cite{limanni2020a}
It bears mentioning however, that Hund's rule on sites $B$ and $C$ can
    not be expressed with a standard representation in most cases, since local
    spin quantum numbers corresponding to $B$ and $C$ are not defined in this
    basis.
Collinear states with $S_{AB} = \{ 0, 3\}$, marked by squares in Figure
    \ref{fig:gbd-compression}, are an exempt from this problem, since they
    can only follow from the coupling of two equal spins.\footnote{ In
        symmetric group parlance, defining $S_{B}$ and $S_{C}$ for the
        branching diagram CSFs is equivalent to reducing the representation
        $\Gamma^{n}$ with respect to the subgroups $\pi_{n_A}$, $\pi_{n_B}$ and
        $\pi_{n_C}$, $n_A + n_B + n_C = n$, instead of the canonical chain
        $\pi_{n - 1}$, $\ldots$, $\pi_1$.\cite{kaplan2013} By inspection of the
        corresponding Young tableaux one can verify that out of all
        non-vanishing CSFs in the right of Figure \ref{fig:gbd-compression}
        only $\ket{uuu\text{ }uuu\text{ }ddd}$ and $\ket{uuu\text{ }ddd\text{
        }uuu}$ have this property, i.e. the boxes labelled by (1,2,3), (4,5,6)
    and (7,8,9) form standard sub-tableaux. }

To mimic the local spin operators of the model, electrons on each site should be
    pre-coupled before combining them into a total spin:
    \begin{equation}
        \ket{L} = \ket{\bar{S}_1^1 \ldots \bar{S}_1^{\text{A}}
            S_{\text{A+1}}^{\text{A+2}} \ldots S_{\text{A+1}}^{\text{A+B}} \bar{S}_{1}^{\text{A+B}}
            \ldots \bar{S}_1^n}, \quad S_A \equiv \bar{S}_{1}^{A}, \quad
            S_B \equiv S_{A+1}^{A+B}, \quad S_{AB} \equiv \bar{S}_{1}^{A+B}.
    \end{equation}
Bars are used to distinguish quantum numbers occurring in both the standard and
    nonstandard representation.

$\ket{C}$ and $\ket{L}$ constitute different coupling schemes that are related by
    a recoupling transformation:
    \begin{equation}
        \ket{L} = \sum_{\ket{C}} \ket{C} \braket{C | L},
        \label{eq:transform-cum-loc}
    \end{equation}
    which we express with the graphical method of spin
    algebra\cite{jucys1962,varshalovich1988,balcar2009} in the style of Wormer
    and Paldus.\cite{wormer2006}
As an aid to the reader, used ingredients of the graphical
    method are briefly recapitulated in Appendix A; a comprehensive
    introduction can be found in reference \citenum{wormer2006}.
The diagrammatic approach is not new to quantum chemistry and has already been
    applied to the study of Serber spin
    functions\cite{wilson1977,wormer1980,paldus1979,paldus1978}, the optimal
    segmentation of two-electron operators in the GUGA\cite{paldus1980} or
    subduction coefficients for a general system partitioning $U(n = n_1
    + n_2) \supset U(n_1) \times U(n_2)$\cite{zhenyi1986,zhenyi1993}
    among others.
Compared to previous work,\cite{zhenyi1986,zhenyi1993} our derivation of the
    recoupling expressions is entirely inductive and does not rely on recursion.
Although less general, this approach leads to simple closed formulas for
    systems with equal spins and we expect them to be convenient
    for studying other model Hamiltonians as well.
Following the introductory example, the recoupling of two sites with three
    electrons into a local spin adapted representation, $S(6) \supset S(3) \times
    S(3)$, will serve as the base case, before treating the general two site
    problem $S(2n) \supset S(n) \times S(n)$ and the chain $S(kn) \supset S(n) \times
    (\ldots)_{k-2} \times S(n)$.
Afterwards, we illustrate with
    $\ce{[CaMn3^{(IV)}O4]}$ and $\ce{[Fe^{(III)}4S4]}$
    cubanes how to use these expressions in practice.

Before $\braket{C|L}$ is amenable to graphical manipulation, it has to be converted
    into a generalised Clebsch-Gordan coefficient.
Inserting the resolution of the identity over all uncoupled states into
    equation \eqref{eq:transform-cum-loc}, we obtain:
\begin{equation}
    \ket{L} = \sum_{\ket{C}}\ket{C}\braket{C|L} =
    \sum_{\ket{C}} \sum_{m_{s_1} \ldots m_{s_i}} \ket{C}\braket{C|s_1 m_{s_1} \ldots
    s_i m_{s_i}}\braket{s_1 m_{s_1} \ldots s_i m_{s_i}|L}
    \label{eq:recoupling}
\end{equation}
    or graphically for $S(6) \supset S(3) \times S(3)$:
\begin{equation}
    % \braket{C|L} =
    \sum_{m_{s_1} \ldots m_{s_6}}
    \begin{tikzpicture}[node distance = \dis, on grid, baseline=(current bounding box.center)]
        \def\dis{1.2cm}
        \node[cgvertex, shape border rotate = 45] (below1) [label=below:$-$]                  {};
        \node[cgvertex, shape border rotate = 45] (below2) [right=of below1, label=below:$-$] {};
        \node[cgvertex, shape border rotate = 45] (below3) [right=of below2, label=below:$-$] {};
        \node[cgvertex, shape border rotate = 45] (below4) [right=of below3, label=below:$-$] {};
        \node[cgvertex, shape border rotate = 45] (below5) [right=of below4, label=below:$-$] {};

        \node[cgvertex, shape border rotate = 135] (above1) [above=3*\dis of below1, label=above:$+$] {};
        \node[cgvertex, shape border rotate = 135] (above2) [right= of above1, label=above:$+$]       {};
        \node[cgvertex, shape border rotate = 135] (above3) [above=3*\dis of below5, label=above:$+$] {};
        \node[cgvertex, shape border rotate = 135] (above4) [below=of above3, label=right:$+$]        {};
        \node[cgvertex, shape border rotate = 135] (above5) [left=of  above4, label=above:$+$]        {};

        \draw (below5) to ["$S_1^5$" below] (below4) to ["$S_1^4$" below] (below3) to ["$S_1^3$" below] (below2) to ["$S_1^2$" below] (below1);
        \draw (above5) to ["$S_4^5$"] (above4) to ["$S_4^6$" right] (above3) to ["$\bar{S}_1^3$" above] (above2) to ["$\bar{S}_1^2$" above] (above1);

        \draw [doublearrow] ($(above5) + (-1,0)$) to ["$s_4$"] (above5);
        \draw [doublearrow] ($(above5) + (0,-1)$) to ["$s_5$"] (above5);
        \draw [doublearrow] ($(above1) + (0,-1)$) to ["$s_2$"] (above1);
        \draw [doublearrow] ($(above1) + (-1,0)$) to ["$s_1$"] (above1);
        \draw [doublearrow] ($(above2) + (0,-1)$) to ["$s_3$"] (above2);
        \draw [doublearrow] ($(above4) + (0,-1)$) to ["$s_6$"] (above4);
        \draw [singlearrow] (above3) to ["$\bar{S}_1^6$" above] ($(above3) + (1, 0)$);

        \draw [singlearrow] (below1) to ["$s_1$"]                     ($(below1) + (-1,0)$);
        \draw [singlearrow] (below1) to ["$s_2$"]                     ($(below1) + (0,1)$) ;
        \draw [singlearrow] (below2) to ["$s_3$"]                     ($(below2) + (0,1)$) ;
        \draw [singlearrow] (below3) to ["$s_4$"]                     ($(below3) + (0,1)$) ;
        \draw [singlearrow] (below4) to ["$s_5$"]                     ($(below4) + (0,1)$) ;
        \draw [singlearrow] (below5) to ["$s_6$"]                     ($(below5) + (0,1)$) ;
        \draw [doublearrow] ($(below5) + (1,0)$) to ["$S_1^6$" below] (below5);
    \end{tikzpicture}
\end{equation}
Contracting over all magnetic quantum numbers and transitioning into 3-$jm$
    symbols yields:
\begin{equation}
  \begin{tikzpicture}[node distance = \dis, on grid, baseline=(current bounding box.center)]
        \def\dis{1.2cm}
        \node[vertex] (below1) [label=below:$+$] {};
        \node[vertex] (below2) [right=of below1, label=below:$+$] {};
        \node[vertex] (below3) [right=of below2, label=below:$+$] {};
        \node[vertex] (below4) [right=of below3, label=below:$+$] {};
        \node[vertex] (below5) [right=of below4, label=below:$+$] {};

        \node[vertex] (above1) [above=2*\dis of below1, label=above:$-$]   {};
        \node[vertex] (above2) [right=of above1, label=above:$-$]          {};
        \node[vertex] (above3) [right=2.5*\dis of above2, label=above:$-$] {};
        \node[vertex] (above4) [below=of above3, label=right:$-$]          {};
        \node[vertex] (above5) [left=of  above4, label=left:$-$]           {};

        \draw [singlearrow] (below2) to ["$S_1^2$" below] (below1);
        \draw [singlearrow] (below3) to ["$S_1^3$" below] (below2);
        \draw [singlearrow] (below4) to ["$S_1^4$" below] (below3);
        \draw [singlearrow] (below5) to ["$S_1^5$" below] (below4);

        \draw [singlearrow] (below1) to ["$s_1$", bend left=60]         (above1);
        \draw [singlearrow] (below1) to ["$s_2$"]                       (above1);
        \draw [singlearrow] (above3) to ["$S_1^6$" right, bend left=80] (below5);

        \draw [singlearrow] (below2) to ["$s_3$"]               (above2);
        \draw [singlearrow] (above1) to ["$\bar{S}_1^2$" above] (above2);
        \draw [singlearrow] (above2) to ["$\bar{S}_1^3$" above] (above3);
        \draw [singlearrow] (above4) to ["$S_4^6$" right]       (above3);
        \draw [singlearrow] (above5) to ["$S_4^5$" above]       (above4);
        \draw [singlearrow] (below5) to ["$s_6$"]               (above4);
        \draw [singlearrow] (below3) to ["$s_4$" left]          (above5);
        \draw [singlearrow] (below4) to ["$s_5$"]               (above5);

        \node[anchor=north] at (current bounding box.south){
            $\delta(S_1^6, \bar{S}_1^6)[S_1^2 S_1^3 S_1^4 S_1^5 ;
                              \bar{S}_1^2 \bar{S}_1^3 S_4^5 S_4^6 ]^{1/2}$
            };

        \node[anchor=north, align=left] at (current bounding box.south){
            $(-1)^{2(s_1 + \bar{S}_1^2 + \bar{S}_1^3 + S_4^5 + s_4) + 2(s_2 + s_3 + s_4 + s_5 + s_6)}$\\
            $= (-1)^{2(s_1 + s_2 + \bar{S}_1^2) + 2(s_4 + s_5 + S_4^5) + 2(\bar{S}_1^3 + s_3 + s_4 + s_6)}
            = (-1)^{2(\bar{S}_1^3 + s_3 + s_4 + s_6)}$
        };
    \end{tikzpicture}
\end{equation}
    where the notation $[S_1 S_2 \ldots] = (2S_1 + 1)(2S_2 + 1)\ldots$ was
    introduced.
To simplify the phase, we make use of the fact that two times the sum of
    angular momenta connected to the same vertex is an even number.
After a two line separation over $S_1^3$/$\bar{S}_1^3$, the arrow
    directions on $\bar{S}_1^2$ and $S_4^5$ are reversed:
\begin{equation}
    \begin{tikzpicture}[node distance = \dis, on grid, baseline=(current bounding box.center)]
        \def\dis{1.3cm}
        \node[vertex] (below1) [label=below:$+$] {};
        \node[vertex] (below2) [right=of below1, label=below:$+$] {};
        \node[vertex] (below3) [right=2*\dis of below2, label=below:$+$] {};
        \node[vertex] (below4) [right=of below3, label=below:$+$] {};
        \node[vertex] (below5) [right=of below4, label=below:$+$] {};

        \node[vertex] (above1) [above=2*\dis of below1, label=above:$-$] {};
        \node[vertex] (above2) [above=2*\dis of below2, label=above:$-$] {};
        % \node[vertex] (above3) [above=2*\dis of below5, label={above:$-$}] {};
        \node[vertex] (above3) [right=3.5*\dis of above2, label=above:$-$] {};
        \node[vertex] (above4) [below=of above3, label={right:$-$}] {};
        \node[vertex] (above5) [left=of  above4, label={above:$-$}] {};

        \draw [singlearrow] (below2) to ["$S_1^2$" below] (below1);
        \draw [singlearrow] (below4) to ["$S_1^4$" below] (below3);
        \draw [singlearrow] (below5) to ["$S_1^5$" below] (below4);

        \draw [singlearrow] (below1) to ["$s_1$", bend left=60]         (above1);
        \draw [singlearrow] (below1) to ["$s_2$"]                       (above1);
        \draw [singlearrow] (above3) to ["$S_1^6$" right, bend left=60] (below5);

        \draw [singlearrow] (below2) to ["$s_3$"]                            (above2) ;
        \draw [singlearrow] (below2) to ["$S_1^3$" right, bend right=60]     (above2) ;
        \draw [singlearrow] (above3) to ["$S_1^3$" left, bend right=60]      (below3) ;
        \draw [singlearrow,red] (above2) to ["$\color{red}{\bar{S}_1^2}$" above] (above1) ;
        \draw [singlearrow] (above4) to ["$S_4^6$" right]                    (above3) ;
        \draw [singlearrow,red] (above4) to ["$\color{red}{S_4^5}$" above]       (above5) ;
        \draw [singlearrow] (below5) to ["$s_6$" right]                      (above4) ;
        \draw [singlearrow] (below3) to ["$s_4$" right]                      (above5) ;
        \draw [singlearrow] (below4) to ["$s_5$" right]                      (above5) ;

        \node[anchor=north] at (current bounding box.south){
            $\delta(S_1^3, \bar{S}_1^3)\delta(S_1^6, \bar{S}_1^6) [S_1^2 S_1^4 S_1^5 ;
                              \bar{S}_1^2 S_4^5 S_4^6 ]^{1/2}$};
        \node[anchor=north] at (current bounding box.south){$
            (-1)^{2(S_1^3 + s_3 + s_4 + s_6)} \color{red}{(-1)^{2\bar{S}_1^2} (-1)^{2S_4^5}}
        $};
    \end{tikzpicture}
    \label{eq:two-sites-s3}
\end{equation}
Following another two line separation over $S_1^2$/$\bar{S}_1^2$ and a three line
    separation over $S_4^5$/$S_1^3$/$S_1^5$, the $6j$ symbols are brought into
    standard form, by changing the vertices and arrows marked in red:
\begin{equation}
    \begin{tikzpicture}[node distance = \dis, on grid, baseline=(current bounding box.center)]
        \def\dis{1.3}
        \node[vertex] (below1) [label=below:$+$] {};
        \node[vertex] (below2) [right=2*\dis of below1, label=below:$+$] {};

        \node[vertex] (above1) [above=2*\dis of below1, label=above:$-$] {};
        \node[vertex] (above2) [above=2*\dis of below2, label=above:$-$] {};

        \draw [singlearrow] (below1) to ["$s_1$", bend left=60]    (above1);
        \draw [singlearrow] (below1) to ["$s_2$"]                  (above1);
        \draw [singlearrow] (below1) to ["$S_1^2$", bend right=60] (above1);

        \draw [singlearrow] (below2) to ["$S_1^2$", bend left=60]  (above2);
        \draw [singlearrow] (below2) to ["$s_3$"]                  (above2);
        \draw [singlearrow] (below2) to ["$S_1^3$", bend right=60] (above2);

        \node[vertex] (below3)  [right=1.5*\dis of below2, label={[red]below:$-$}] {};
        \node[vertex] (below4)  [right=2*\dis of below3, label={[red]below:$-$}] {};
        \node[vertex] (high1)   [right=2.5*\dis of above2, label=above:$-$] {};
        \node[vertex] (middle1) [below=1.2*\dis of high1, label={left:$-$}] {};
        % \node[vertex] (middle1) [above right=0.8cm and 1cm of below3, label={left:$-$}] {};
        % \node[vertex] (high1)   [above=1.2cm of middle1, label=above:$-$] {};
        \draw [singlearrow] (below4) to ["$S_1^4$" below]                           (below3) ;
        \draw [singlearrow,red] (below3)  to ["$\color{red}{S_1^3}$" left, bend left=30](high1) ;
        \draw [singlearrow] (high1)  to ["$S_1^5$" right, bend left=30]                   (below4) ;
        \draw [singlearrow] (below3) to ["$s_4$" below]                             (middle1);
        \draw [singlearrow] (below4) to ["$s_5$" below]                             (middle1);
        \draw [singlearrow] (high1)  to ["$S_4^5$"]                                 (middle1);

        \node[vertex] (below5) [right= of below4, label={[red]below:$-$}] {};
        \node[vertex] (below6) [right=2*\dis of below5, label={[red]below:$-$}] {};
        \node[vertex] (high2)  [right=3*\dis of high1, label=above:$-$] {};
        \node[vertex] (middle2)[below=1.2*\dis of high2, label={left:$-$}] {};
        % \node[vertex] (middle2) [above right=0.8cm and 1cm of below5, label=right:$-$] {};
        % \node[vertex] (high2) [above=1.2cm of middle2, label=above:$-$] {};
        \draw [singlearrow,red] (below5)   to ["$\color{red}{S_1^3}$" left, bend left=30](high2) ;
        \draw [singlearrow] (below6)  to ["$S_1^5$" below]                            (below5) ;
        \draw [singlearrow] (high2)   to ["$S_1^6$" right, bend left=30]              (below6) ;
        \draw [singlearrow] (middle2) to ["$S_4^5$" right]                            (below5) ;
        \draw [singlearrow,red] (middle2)  to ["$\color{red}{s_6}$" below]                (below6);
        \draw [singlearrow] (middle2) to ["$S_4^6$" right]                            (high2);

        \node[anchor=north] at (current bounding box.south){
          $\delta(S_1^2, \bar{S}_1^2)\delta(S_1^3, \bar{S}_1^3)\delta(S_1^6, \bar{S}_1^6) [S_1^4 S_1^5 ; S_4^5 S_4^6 ]^{1/2}$
        };
        \node[anchor=north, align=left] at (current bounding box.south){
          $(-1)^{2(S_1^3 + s_3 + s_4 + s_6) + 2S_1^2 + 2S_4^5} \times$\\
                $~~~~~\times \color{red}{(-1)^{4S_1^3}}
                \color{red}{(-1)^{2s_6}}
                \color{red}{(-1)^{S_1^3 + s_4 + S_1^4}}
                \color{red}{(-1)^{S_1^4 + s_5 + S_1^5}}
                \color{red}{(-1)^{S_1^3 + S_4^5 + S_1^5}}
                \color{red}{(-1)^{S_1^5 + s_6 + S_1^6}}$\\
            $=(-1)^{2(S_1^3 + s_3 + s_4 + s_6) + 2S_1^2 + 3S_4^5 +
                2s_6 + 2S_1^3 + s_4 + 2S_1^4 + s_5 + 3S_1^5 + s_6 + S_1^6}$\\
            $=(-1)^{(2s_3 + 2s_4 + 2S_1^2 + 2S_1^4) + 3S_4^5 + s_4 + s_5 + 3S_1^5 + s_6 + S_1^6}$\\
            $=(-1)^{3S_4^5 + s_4 + s_5 + 3S_1^5 + s_6 + S_1^6} = (-1)^{(S_1^3 + S_1^5 + s_4 + s_5) + (S_1^3 + s_6 + S_4^5 + S_1^6)}$
            };
    \end{tikzpicture}
\end{equation}
    We also used twice an identity for momenta connected to the same vertex,
    $(-1)^{2S_{12} + 2S_3} = (-1)^{2S_{13}}$ .
The genealogical CSFs $\ket{C}$ fulfil the triangle
    condition for cumulative spins by construction and the $3j$ symbols take
    the value one accordingly.
Substituting $s_i = \frac{1}{2}$, using the symmetry of the $6j$ symbol
    under the column permutation $(1 2 3) \leftrightarrow (3 2 1)$ and flipping
    columns one and three, the recoupling coefficients
    $\braket{C|L}$ for $S(6) \supset S(3) \times S(3)$ take the form:
\begin{equation}
    \begin{split}
    &\delta(S_1^2, \bar{S}_1^2)\delta(S_1^3, \bar{S}_1^3)\delta(S_1^6,
    \bar{S}_1^6) \\
    &\times (-1)^{(S_1^3 + S_1^5 + 1) + (S_1^3 + \frac{1}{2} + S_4^5 + S_1^6)}
    \left[ S_1^4 S_4^5 ; S_1^5 S_4^6 \right]^{1/2}
    \begin{Bmatrix} S_1^5 & \frac{1}{2} & S_1^4 \\[2pt] \frac{1}{2} & S_1^3 & S_4^5 \end{Bmatrix}
    \begin{Bmatrix} S_1^6 & \frac{1}{2} & S_1^5 \\[2pt] S_4^5       & S_1^3 & S_4^6 \end{Bmatrix}
    \end{split}.
    \label{eq:s6-recoupling}
\end{equation}
If we introduce the weighted symbol $\widetilde{W}$ related to the original
    Racah coefficient $\overline{W}$\cite{racah1942}:
\begin{equation}
  \begin{aligned}
        \widetilde{W} \begin{pmatrix} j_1 & j_2 & J_{12} \\ j_3 & J & J_{23} \end{pmatrix} &=
        [J_{12} J_{23}]^{1/2} \overline{W} \begin{pmatrix} j_1 & j_2 & J_{12} \\ j_3 & J & J_{23} \end{pmatrix}
        \\ &=
        [J_{12} J_{23}]^{1/2} (-1)^{j_1 + j_2 + j_3 + J} \begin{Bmatrix} j_1 &
        j_2 & J_{12} \\ j_3 & J & J_{23} \end{Bmatrix},
  \end{aligned}
  \label{eq:s6-racah}
\end{equation}
    we can write equation \eqref{eq:s6-recoupling} compactly as:
\begin{equation}
    \delta(S_1^2, \bar{S}_1^2)\delta(S_1^3, \bar{S}_1^3)\delta(S_1^6, \bar{S}_1^6) \text{ }
    \widetilde{W} \begin{pmatrix} S_1^5 & \frac{1}{2} & S_1^4 \\[2pt] \frac{1}{2} & S_1^3 & S_4^5 \end{pmatrix}
    \widetilde{W} \begin{pmatrix} S_1^6 & \frac{1}{2} & S_1^5 \\[2pt] S_4^5       & S_1^3 & S_4^6 \end{pmatrix}.
    \label{eq:s6-recoupling-compact}
\end{equation}
Equation \eqref{eq:s6-recoupling-compact} may be generalised to the case of $k$ sites
    with $S_i = \frac{3}{2}$.
Due to the local spin coupling scheme, different sites remain separable over two
    lines, entirely similar to $S_{1}^{3}$ / $\bar{S}_{1}^{3}$ in equation
    \eqref{eq:s6-recoupling} and introduce a factor analogous to
    \eqref{eq:s6-racah} with a constant offset:
\begin{equation}
    \begin{split}
        \delta(S_1^2, \bar{S}_1^2) &\delta(S_1^{3k}, \bar{S}_1^{3k}) \prod_{i=0}^{k - 2}
    \delta(S_1^{3+3i}, \bar{S}_1^{3+3i}) \\
    &\widetilde{W}\begin{pmatrix}
        S_1^{5+3i}       & \frac{1}{2}      & S_1^{4+3i} \\[2pt]
        \frac{1}{2}      & S_1^{3+3i}       & S_{4+3i}^{5+3i}
    \end{pmatrix}
    \widetilde{W}\begin{pmatrix}
        S_1^{6+3i}       & \frac{1}{2}      & S_1^{5+3i} \\[2pt]
        S_{4+3i}^{5+3i}  & S_1^{3+3i}       & S_{4+3i}^{6+3i}
    \end{pmatrix}.
    \label{eq:chain-recoupling}
    \end{split}
\end{equation}
To generalise to arbitrary spins $S(n)$, it is instructive to first observe the action
    of adding one electron to each of two sites, as in $S_8 \supset S_4 \times S_4$
    (phases and factors omitted):
\begin{equation}
    \begin{tikzpicture}[scale=1.4, baseline=(current bounding box.center)]
        \def\dis{1.3}
        \node[vertex] (below1a) [label=below:$+$] {};
        \node[vertex] (below2a) [right=of below1a, label=below:$+$] {};
        \node[vertex] (below3a) [right=of below2a, label=below:$+$] {};

        \node[vertex] (above1a) [above=2*\dis of below1a, label=above:$-$] {};
        \node[vertex] (above2a) [above=2*\dis of below2a, label=above:$-$] {};
        \node[vertex] (above3a) [above=2*\dis of below3a, label=above:$-$] {};

        \draw [singlearrow] (below2a) to ["$S_1^2$" below] (below1a);
        \draw [singlearrow] (below3a) to ["$S_1^3$" below] (below2a);
        \draw [singlearrow] (above1a) to ["$\bar{S}_1^2$" above] (above2a);
        \draw [singlearrow] (above2a) to ["$\bar{S}_1^3$" above] (above3a);
        \draw [singlearrow] (below1a) to ["$s_1$", bend left=60]  (above1a);
        \draw [singlearrow] (below1a) to ["$s_2$"]                (above1a);
        \draw [singlearrow] (below2a) to ["$s_3$"]                (above2a) ;
        \draw [singlearrow] (below3a) to ["$s_4$"]                (above3a) ;
        \draw [singlearrow] (below3a) to ["$S_1^4$" right, bend right=60]     (above3a) ;
    \end{tikzpicture}
    \begin{tikzpicture}[scale=1.4, baseline=(current bounding box.center)]
        \def\dis{1.3}
        \node[vertex] (below1) [label=below:$+$] {};
        \node[vertex] (below2) [right=1.2*\dis of below1, label=below:$+$] {};
        \node[vertex] (below3) [right=0.7*\dis of below2, label=below:$+$] {};
        \node[vertex] (below4) [right=1.33*\dis of below3, label=below:$+$] {};

        % \node[vertex,white] (above0) [above=2*\dis of below1] {};
        \node[vertex] (above3) [above=\dis of below3, label=above:$-$] {};
        \node[vertex] (above2) [right= of above3, label=right:$-$] {};
        \node[vertex] (above4) [left=1.33*\dis of above3, label=above:$-$] {};
        \node[vertex] (above1) [above=\dis of above2, label=above:$-$] {};

        \draw [singlearrow] (above1) to ["$S_1^8$" right, bend left=60]  (below4);
        \draw [singlearrow] (above1) to ["$S_1^4$" left=0.25cm, bend right=55] (below1);
        \draw [singlearrow] (above2) to ["$S_4^8$" right]                (above1);
        \draw [singlearrow] (above3) to ["$S_4^7$" above]                (above2);
        \draw [singlearrow] (above4) to ["$S_4^6$" above]                (above3);

        \draw [singlearrow] (below4) to ["$s_8$" right]       (above2);
        \draw [singlearrow] (below3) to ["$s_7$" right]       (above3);
        \draw [singlearrow] (below2) to ["$s_6$" right]       (above4);
        \draw [singlearrow] (below1) to ["$s_5$" right] (above4);

        \draw [singlearrow] (below2) to ["$S_1^5$" below] (below1);
        \draw [singlearrow] (below3) to ["$S_1^6$" below] (below2);
        \draw [singlearrow] (below4) to ["$S_1^7$" below] (below3);
    \end{tikzpicture}
\end{equation}
Compared to \eqref {eq:two-sites-s3}, an additional $3j$ symbol can be factored
    out from the left drawing and two extra vertices are formed on the right.
After separating over three lines twice, $S_1^4 / S_4^6 / S_1^6$ and $S_1^4 / S_4^7 /
    S_1^7$, the right hand side reduces to a product of three $6j$ symbols, i.e. every
    additional electron contributes one $6j$ coefficient per site.
Taking the phases into account as previously, the recoupling coefficient for
    $S(2n) \supset S(n) \times S(n)$ becomes a product of $\widetilde{W}$ over site electrons $n$:
\begin{equation}
    \prod_{p = 2}^{n - 1} \delta(S_1^{p}, \bar{S}_1^{p}) \times
    \delta(S_1^{n}, \bar{S}_1^{n})
    \delta(S_1^{2n}, \bar{S}_1^{2n})
    \prod_{m = 0}^{n - 2}
    \widetilde{W}
    \begin{pmatrix}
        S_1^{2 + m + n}       & \frac{1}{2} & S_1^{1 + m + n} \\[2pt]
        S_{1 + n}^{1 + m + n} & S_1^n       & S_{1 + n}^{2 + m + n}
    \end{pmatrix}.
    \label{eq:s2n-recoupling}
\end{equation}
The general expression for recoupling a chain of equivalent spins $S_i \geq 1$
    into $S(kn) \supset S(n) \times (\ldots)_{k-2} \times S(n)$ is a combination of
    equations\,\eqref{eq:chain-recoupling} and \eqref{eq:s2n-recoupling}.
Every site adds another $2n$ vertices to the graph, yielding
    $(n - 1)$ $6j$ symbols.
In total the recoupling coefficient is a product of $(k - 1)(n - 1)$ Racah
    coefficients with the corresponding Kronecker deltas:
\begin{equation}
    \begin{split}
    \braket{C|L} =
    &\prod_{p = 2}^{n - 1} \delta(S_1^{p}, \bar{S}_1^{p}) \times
        \delta(S_1^{ k n }, \bar{S}_1^{ k n })
        \\
    &\times \prod_{i = 0}^{k - 2} \prod_{m = 0}^{n - 2}
    \delta(S_1^{ n(i+1) }, \bar{S}_1^{ n(i+1) })
    \widetilde{W}\begin{pmatrix}
        S_1^{2 + m + n(i+1)}            & \frac{1}{2}  & S_1^{1 + m + n(i+1)} \\[2pt]
        S_{1 + n(i+1)}^{1 + m + n(i+1)} & S_1^{n(i+1)} & S_{1 + n(i+1)}^{2 + m + n(i+1)}
    \end{pmatrix}.
    \end{split}
    \label{eq:sn-chain-recoupling}
\end{equation}

\section{Application: \ce{[CaMn^{(IV)}3O4]} Cubane}
\begin{figure}[H]
    \includegraphics[width=0.4\textwidth]{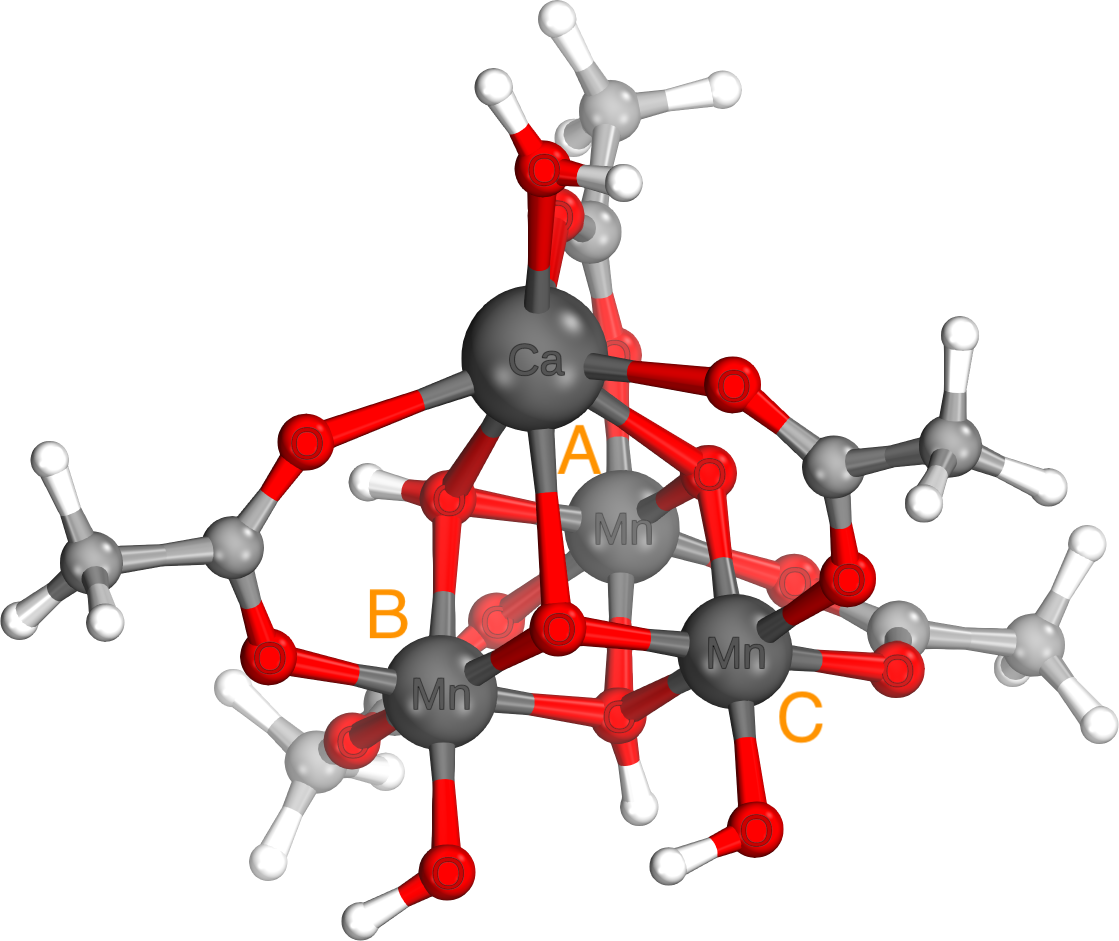}
    \caption{Geometry of the doubly protonated \ce{[CaMn^{(IV)}3O4]} cubane
        model.\cite{krewald2013} Magnetic sites are labelled $A, B, C$ following equation
        \eqref{eq:heisenberg-general}. The picture was created with
        IboView.\cite{knizia2013, knizia2015}}
    \label{fig:camn_cubane}
\end{figure}

The \ce{[CaMn^{(IV)}3O4]} cubane shown in Figure \ref{fig:camn_cubane} is
    inspired by a synthetic model of the oxygen evolving
    complex of photosystem II.
Compared to the experimental structure,\cite{mukherjee2012} the ligands were
    simplified and the bridging oxygen in the \ce{Mn_A/Mn_B} plane
    protonated,\cite{krewald2013} to study the impact of oxo-protonation on the
    electronic structure.
In the original study on this compound,\cite{krewald2013} broken-symmetry (BS) DFT
    calculations assuming an isotropic, three-parameter Heisenberg model:
\begin{equation}
    \hat{\mathcal{H}}_{3J} = \sum_{i < j}^{\{ A,B,C \}} J_{ij} (\hat{S}_i
    \cdot \hat{S}_j), %+ K_{ij} (\hat{S}_i \cdot \hat{S}_j)^2
    \label{eq:heisenberg-general}
\end{equation}
    predicted an octet ground state with couplings constants $J_{AB} =
    \SI{6.2}{\per\cm}$, $J_{AC} = \SI{-40.2}{\per\cm}$ and $J_{BC} =
    \SI{-25.4}{\per\cm}$.
Experimental estimates for this structure are not available.
With three distinct couplings, equation \eqref{eq:heisenberg-general} is no
    longer analytically solvable, making it an interesting target for the effective
    Hamiltonian approach.
Another advantage of the scheme is a straightforward
    incorporation of higher-order interactions. In the following we
    consider a more general expression including biquadratic interactions $K_{ij}
    (\hat{S}_i \cdot \hat{S}_j)^2$, denoted $\hat{\mathcal{H}}_{3J3K}$, see
    also section S3 in the supporting information.

Apart from the extra recoupling step described in the previous section,
    formation of the effective Hamiltonian follows the standard
    prescription,\cite{degraaf2016,malrieu2013,maurice2009} i.e. (i) obtain the CI
    vector in the standard CSF basis, (ii) transform relevant CSFs into the
    required coupling scheme, (iii) project onto the model space, ensuring that the
    discarded norm is small, (iv) orthonormalise the projections and (v) invert the
    spectral decomposition with the ab initio energies.
Magnetic couplings can then be obtained from a mean-square fit of the
    numerical to the analytical matrix.
In the first step, the phase convention of the CI vector should be accounted
    for.
Recoupling CSFs in the unitary group formalism, as for instance used in
    \texttt{OpenMolcas},\cite{limanni2023b} requires a phase to be
    applied beforehand:\cite{paldus2020b,drake1977}
    \begin{equation}
        \phi = (-1)^{2\left(\sum_{k \in \{l_{23}\}} S_1^{k - 1}\right)},
    \end{equation}
    where $\{l_{23}\}$ denotes level indices with values 2 or 3 (negative
    coupling or doubly occupied) in UGA $\mathbf{d}$ vector notation.\cite{paldus2020b}
In line with the introductory example, the $S_{\text{tot}} = \frac{3}{2}$ block
    of the model matrix is shown in Table \ref{tab:model-quartet}.
The remaining blocks are given in the supporting information.
\begin{table}[H]
    \caption{Analytical matrix corresponding to $\hat{\mathcal{H}}_{3J3K}$
        in the basis $\ket{S_{tot} = \frac{3}{2},
        S_{AB}}$. Zero matrix elements were omitted for clarity. We made the substitutions
        $J_{3m} = J_{13} - J_{23}$, $J_{3p} = J_{13} + J_{23}$ and analogously
        for $K_{3m}$ and $K_{3p}$, to make clear that setting $J_{13} = J_{23}$ and
        $K_{ij} = 0$ diagonalises the model.\cite{kambe1950,griffith1972a}}
    \begin{tabular}{@{}r cc@{}}
        $\hat{\mathcal{H}}_{3J3K}$ & $\ket{\frac{3}{2}, 0}$ & $\ket{\frac{3}{2}, 1}$ \\
        \cmidrule{1-3}
        $\bra{\frac{3}{2}, 0}$ & $-\frac{15}{16} (4 J_{12}+5 (3 K_{12} + K_{3p}))$ & $-\frac{5}{8} \sqrt{3} (2 J_{3m}+K_{3m})$                      \\
        $\bra{\frac{3}{2}, 1}$ & $-\frac{5}{8} \sqrt{3} (2 J_{3m}+K_{3m})$         & $\frac{1}{80} (-5 (44 J_{12}+8 J_{3p}+121 K_{12})-587 K_{3p})$ \\
        $\bra{\frac{3}{2}, 2}$ & $-\frac{3 \sqrt{5}}{2}K_{3p}$                     & $-2 \sqrt{\frac{3}{5}} (J_{3m}+2 K_{3m})$                      \\
        $\bra{\frac{3}{2}, 3}$ &                                                   & $-\frac{3 \sqrt{21}}{10}    K_{3p}  $                          \\
      \\[-0.5em]
      $\hat{\mathcal{H}}_{3J3K}$ & $\ket{\frac{3}{2}, 2}$ & $\ket{\frac{3}{2}, 3}$ \\
        \cmidrule{1-3}
        $\bra{\frac{3}{2}, 0}$ & $-\frac{3 \sqrt{5} }{2}K_{3p}$                         &                                                                   \\
        $\bra{\frac{3}{2}, 1}$ & $-2 \sqrt{\frac{3}{5}} (J_{3m}+2 K_{3m})$              & $-\frac{3 \sqrt{21} }{10}K_{3p}$                                  \\
        $\bra{\frac{3}{2}, 2}$ & $-\frac{3}{16} (4 J_{12}+8 J_{3p}+3 K_{12}+29 K_{3p})$ & $-\frac{3}{8} \sqrt{\frac{7}{5}} (2 J_{3m}+9 K_{3m})$             \\
        $\bra{\frac{3}{2}, 3}$ & $-\frac{3}{8} \sqrt{\frac{7}{5}} (2 J_{3m}+9 K_{3m})$  & $\frac{9 J_{12}}{4}-3 J_{3p}-\frac{27}{80} (15 K_{12}+29 K_{3p})$ \\
    \end{tabular}
    \label{tab:model-quartet}
\end{table}
The corresponding numerical matrices were derived from calculations
    performed with \texttt{OpenMolcas}\cite{limanni2023b} using the ANO-RCC basis
    set\cite{roos2003} with contractions
    Ca(20s16p)/[5s4p], Mn(21s15p10d)/[5s4p2d], O(14s9p)/[3s2p],
    C(14s9p)/[2s1p], H(8s)/[1s] and the Cholesky decomposition of two-electron
    integrals with a threshold of $10^{-4}$.\cite{pedersen2009,aquilante2011}

\sloppypar{
A minimal description of the magnetic interactions is achieved by correlating
    the nine electrons occupying the $t_{2g}$ orbitals in a CAS(9,9).
Spin models usually assume a common spatial component, hence state-averaged
    CASSCF over the ground state of every multiplicity in equation
    \eqref{eq:cg-trinuclear} with equal weight was performed.\footnote{Possible in
        \texttt{OpenMolcas} through a modification of the \texttt{\&RASSCF}/FCIQMC
    interface.}
The $S_{\text{tot}} = \frac{3}{2}$ block of the numerical matrix is shown in Table
    \ref{tab:numerical-minimal}; all other blocks can be found in the supporting
    information.}
\begin{table}[H]
    \caption{Effective Hamiltonian obtained from CASSCF(9,9) with labels
        $\ket{S_{\text{tot}} = \frac{3}{2}, S_{AB}}$ relative to $E(S_{\text{tot}} =
    \frac{9}{2})$ in units of $\SI{}{\per\cm}$. Zeros were omitted to enhance legibility.}
    \begin{tabular}{@{}r cccc@{}}
        $\hat{\mathcal{H}}_\mathrm{eff}^\mathrm{CASSCF}$ & $\ket{\frac{3}{2}, 0}$ & $\ket{\frac{3}{2}, 1}$ & $\ket{\frac{3}{2}, 2}$ & $\ket{\frac{3}{2}, 3}$ \\
        \cmidrule{1-5}
        $\bra{\frac{3}{2}, 0}$ & 147 &  17 &     &     \\[2pt]
        $\bra{\frac{3}{2}, 1}$ &  17 & 165 &  12 &     \\[2pt]
        $\bra{\frac{3}{2}, 2}$ &     &  12 & 199 &   7 \\[2pt]
        $\bra{\frac{3}{2}, 3}$ &     &     &   7 & 251 \\
    \end{tabular}
    \label{tab:numerical-minimal}
\end{table}
Interactions constants are all ferromagnetic at this level of theory
    and sufficiently different to warrant a 3$J$ model, see the second row of Table
    \ref{tab:mcpdft-js-ks}.
Due to a mean-field description of virtual ligand-to-metal charge transfer,
    minimal CASSCF magnetic orbitals tend to be too localised on the
    metals.\cite{angeli2012}
Rather than trying to capture this interaction perturbatively, we expand the
    variational space within the restricted active space (RAS) framework to
    keep exact diagonalisation feasible.\cite{malmqvist1990}
Six unprotonated bridging \ce{O} $2p$ and three $\sigma_{\text{Mn-O}}$ were added
    to RAS1 and six empty \ce{Mn} $e_g$ orbitals to RAS3, with up to two holes and
    particles, respectively, which we abbreviate as RAS(27,24).
Orbital pictures are provided in the supporting information.
These active orbitals are expected to make the largest differential contributions to
    the superexchange mechanism and prevent some of the \ce{Mn} $e_g$ to be
    rotated out in favour of \ce{O} $2p'$ orbitals.
Analogous to the smaller CAS(9,9), a state-averaged RASSCF over
    the ground state of each spin sector was performed.
Correlating the ligands introduces delocalisation tails into the metal-centred
    orbitals and reduces the retained wave function norm after projection onto the
    model space to $\approx 78\%$.
This number may appear small, but a reduced weight of the magnetic manifold is
    to be expected with an increasing ligand-to-metal charge transfer in the
    extended active space.
The RAS retains a dectet ground state, but compresses the gaps
    between states, as can be seen for the quartet block in Table
    \ref{tab:numerical-extended}.
All remaining spin sectors are given in the supporting information.
\begin{table}[H]
    \caption{Effective Hamiltonian obtained from RASSCF(27, 24) with labels
        $\ket{S_{\text{tot}} = \frac{3}{2}, S_{AB}}$ relative to $E(S_{\text{tot}} =
    \frac{9}{2})$ in units of $\SI{}{\per\cm}$. Zeros were omitted to enhance legibility.}
    \begin{tabular}{@{}r cccc@{}}
        $\hat{\mathcal{H}}_\mathrm{eff}^\mathrm{RASSCF}$ & $\ket{\frac{3}{2}, 0}$ & $\ket{\frac{3}{2}, 1}$ & $\ket{\frac{3}{2}, 2}$ & $\ket{\frac{3}{2}, 3}$ \\
        \cmidrule{1-5}
        $\bra{\frac{3}{2}, 0}$ & 79 &  40 &     &     \\[2pt]
        $\bra{\frac{3}{2}, 1}$ & 40 &  89 &  28 &     \\[2pt]
        $\bra{\frac{3}{2}, 2}$ &    &  28 & 107 &  16 \\[2pt]
        $\bra{\frac{3}{2}, 3}$ &    &     &  16 & 136 \\
    \end{tabular}
    \label{tab:numerical-extended}
\end{table}
In contrast to the minimal CAS, the coupling constants from RASSCF, listed in
    the third row of Table \ref{tab:mcpdft-js-ks}, indicate an almost symmetric
    interaction between \ce{Mn_A}/\ce{Mn_B} and \ce{Mn_B}/\ce{Mn_C}.
Biquadratic contributions remain negligible, supporting the simplified model
    employed in reference \citenum{krewald2013}.

With an improved reference wave function, it makes sense to also consider dynamic
    correlation effects outside the active space.
Here we settled for multi-configurational pair density functional theory
    (MCPDFT),\cite{carlson2015,limanni2014,lehtola2018} due to its favourable
    cost-to-performance ratio compared to RASPT2.\cite{malmqvist2008}
Notably, the tested functionals consistently predict a quartet ground state with
    purely antiferromagnetic couplings while maintaining symmetric $J_{12} /
    J_{23}$ exchange pathways, as shown in the last three rows of Table
    \ref{tab:mcpdft-js-ks}.
Couplings obtained with the SCAN-E0\cite{sun2015} functional are more than
    twice as large as those from PBE\cite{perdew1996,perdew1997} or
    BLYP\cite{becke1988,lee1988,miehlich1989}, but display the same qualitative
    trend.
Considering excitation level restrictions of the RASSCF and basis set
    limitations, our estimates are unlikely to be converged.
Full CASSCF calculations are required to understand the discrepancy between the
    BS-DFT and RASSCF+MCPDFT results, which will be subject of future work.
\begin{table}[H]
    \caption{Magnetic coupling constants in units of $\SI{}{\per\cm}$ derived
    from CASSCF(9,9) and different fully translated pair density functionals
    with a RASSCF(27, 24) reference wave function, as well as the BS-DFT
    estimate from reference \citenum{krewald2013}. The final two columns show
    the intercept of the mean square fit (in units of $\SI{}{\per\cm}$) and the coefficient of determination.}
    \begin{tabular}{@{}r cccccccc@{}}
                                  & $J_{12}$ & $J_{13}$ & $J_{23}$ & $K_{12}$ & $K_{13}$ & $K_{23}$ & $b$ & $R^2$\\
        \cmidrule{1-9}
        BS-DFT\cite{krewald2013} &  6.2     & -40.2    & -25.4    & -       & -       & -    & -     & -     \\
        CASSCF(9, 9)              & -6.6     & -27.7    & -20.0   & 0.0     & 0.0     & 0.0  &-0.01  &1.0000 \\
        RASSCF(27, 24)            & -3.5     & -22.2    & -3.7    & 0.0     & 0.0     & -0.1 &-0.16  &0.9999 \\
        +MCPDFT(ftPBE)            & 19.8     & 5.8      & 19.8    & -0.1    & 0.3     & -0.3 &-0.14  &0.9998 \\
        +MCPDFT(ftBLYP)           & 21.9     & 7.9      & 21.9    & -0.1    & 0.3     & -0.3 &-0.14  &0.9998 \\
        +MCPDFT(ftSCAN-E0)       & 54.0     & 23.4     & 53.9     & -0.3    & 0.5     & -0.6 &-0.21  &0.9998 \\
    \end{tabular}
    \label{tab:mcpdft-js-ks}
\end{table}

\section{Non-Sequential Coupling Patterns}
\begin{figure}[H]
    \begin{tikzpicture}[node distance = \dis, on grid, baseline=(current bounding box.center)]
        \def\dis{1.5cm}
        \node[vertex] (head) [label=above:$-$] {};
        \node[vertex] (child_AB) [above left=of  head, label=below:$-$] {};
        \node[vertex] (child_CD) [above right=of head, label=below:$-$] {};
        \node         (S_A) [above left=of child_AB]  {};
        \node         (S_B) [above right=of child_AB] {};
        \node         (S_C) [above left=of child_CD]  {};
        \node         (S_D) [above right=of child_CD] {};

        \draw [singlearrow] (child_AB) to ["$S_{AB}$" below left]  (head);
        \draw [singlearrow] (child_CD) to ["$S_{CD}$" below right] (head);
        \draw [singlearrow] (S_A)      to ["$S_A$" below left] (child_AB);
        \draw [singlearrow] (S_B)      to ["$S_B$" above left = 0cm and -1mm] (child_AB);
        \draw [singlearrow] (S_C)      to ["$S_C$" above right = 0cm and -1mm] (child_CD);
        \draw [singlearrow] (S_D)      to ["$S_D$" below right] (child_CD);
        \draw [singlearrow] (head)     to ["$S_{tot}$" right] (0,-1);
    \end{tikzpicture}
    \caption{Coupling of four angular momenta in the $((AB)(CD))$ scheme,
    closely resembling a binary tree.}
    \label{fig:bbt-csf}
\end{figure}
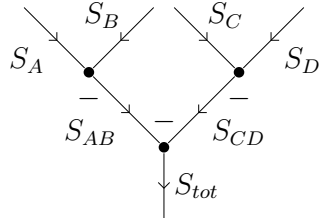

Many transition metal clusters like the oxygen evolving complex
    or FeMo cofactor are formed by four or more magnetic sites.
The sequential coupling of local spins implied by equation
    \eqref{eq:sn-chain-recoupling} will suffice to construct the effective
    Hamiltonian for these compounds, but it is known that certain models
    admit diagonal form in non-sequential orderings.
A relevant example is the four site bilinear Heisenberg model:
\begin{equation}
    \hat{\mathcal{H}}_{4J} = \sum_{i < j}^{\{ A, B, C, D \} } J_{ij} (\hat{S}_{i} \cdot
    \hat{S}_j), \quad J_{AC} = J_{BC} = J_{AD} = J_{BD}.
    \label{eq:4j-heisenberg}
\end{equation}
    which is diagonal in the $((AB)(CD))$ coupling scheme.\cite{griffith1972b}
Drawn as a spin graph, see Figure \ref{fig:bbt-csf}, this arrangement of spins
    resembles a balanced binary
    tree and thus the abbreviation
\begin{equation}
    \ket{T} = \ket{\ldots \bar{S}_1^{\text{A}} \ldots
        S_{\text{A+1}}^{\text{B}} \bar{S}_1^{\text{A+B}}
        \ldots S_{\text{A+B+1}}^{\text{C}} \ldots S_{\text{C+1}}^{\text{D}}
        S_{\text{A+B+1}}^{\text{C+D}} \bar{S}_{\text{1}}^\text{{A+B+C+D}}}
\end{equation}
    will be used for these CSFs.
Local spins $A, B, C, D$ are already defined for $\ket{L}$ CSFs, hence the
    remaining transformation from \eqref{eq:sn-chain-recoupling}, $(((AB)C)D)
    \rightarrow ((AB)(CD))$, can be expressed as a single Racah coefficient:
\begin{equation}
    \braket{L|T} = \widetilde{W}
    \begin{pmatrix}
        S_{\text{AB}} & S_{\text{C}} & S_{\text{ABC}} \\
        S_{\text{D}}& S_{\text{tot}} & S_{\text{CD}}
    \end{pmatrix},
    \label{eq:cube-recoupling}
\end{equation}
    allowing us to examine the properties of the ab initio wave function in
    this basis.
As an example, an all-ferric \ce{[Fe^{(III)}4S4(SMe)4]} cubane model system
    will be considered.\cite{moula2018,limanni2021b}
The singlet manifold of this particular cluster was previously studied with
    GUGA-FCIQMC\cite{limanni2021b}.
To illustrate the concept, we concentrate on the exactly
    diagonalisable $S_{\text{tot}} = 8$ block of the minimal CAS(20,20)
    comprising the iron $3d$ orbitals.
Calculations were performed with \texttt{OpenMolcas}\cite{limanni2023b} using
    the ANO-RCC basis set\cite{roos2003} with
    contractions Fe(21s15p10d)/[5s4p2d], S(17s12p)/[4s3p], C(14s9p)/[2s1p],
    H(8s)/[1s] and the Cholesky decomposition of two-electron integrals with a
    threshold of $10^{-4}$.\cite{pedersen2009,aquilante2011}

The model space of the $S_{\text{tot}} = 8$ manifold is spanned by six states, resulting from the coupling
    of four $S_{i} = \frac{5}{2}$.
From the branching diagram, 170 open-shell CSFs can be identified in a
    localised basis, entirely similar to the left side of Figure \ref{fig:gbd-compression}.
Table \ref{tab:sparsity} lists the $L_1$ norm, $1 \leq \sum_i^{n_{CSF}} |c_i|
    \leq \sqrt{n_{CSF}}$, and inverse participation ratio (IPR),
    $\frac{1}{n_{CSF}} \leq \sum_i^{n_{CSF}} c_i^4 \leq 1$, as measures of
    sparsity in this generic basis $\ket{G}$ and compares it to the
    standard $\ket{C}$, local sequential $\ket{L}$ and binary tree $\ket{T}$
    coupling schemes in the site-ordered basis.
Assuming the worst-case scenario, orbitals in $\ket{G}$ are ordered such that
    electrons one and two are on different atoms, preventing partial fulfilment
    of Hund's rule on the first site.
\begin{table}[H]
    \caption{Sparsity measures for the branching diagram CSFs of the first six
    roots $\ket{S_{tot} = 8}$ in the singly occupied manifold $\ket{\bar{\psi}}$ of the CAS(20,20). Small values of the $L_1$ norm
    and large IPRs indicate sparse wave functions.}
    \label{tab:sparsity}
    \begin{tabular}{@{}r cccccc@{}}
    $S_{\text{tot}} = 8$    & $\ket{\bar{\psi}_1}$ & $\ket{\bar{\psi}_2}$ & $\ket{\bar{\psi}_3}$ & $\ket{\bar{\psi}_4}$ & $\ket{\bar{\psi}_5}$ & $\ket{\bar{\psi}_6}$  \\
    \cmidrule{1-7}
    $L_1$ norm \\
    $\ket{G}$ & 10.4 & 10.3 & 10.3 & 11.4 & 7.86 & 10.4 \\
    $\ket{C}$ & 6.80 & 6.76 & 6.70 & 5.82 & 6.80 & 6.45 \\
    $\ket{L}$ & 1.46 & 1.97 & 1.96 & 1.72 & 1.46 & 1.69 \\
    $\ket{T}$ & 1.07 & 1.48 & 1.48 & 1.06 & 1.06 & 1.04 \\
    IPR \\
    $\ket{G}$ & 0.01 & 0.01 & 0.01 & 0.01 & 0.03 & 0.01 \\
    $\ket{C}$ & 0.03 & 0.04 & 0.04 & 0.04 & 0.03 & 0.03 \\
    $\ket{L}$ & 0.56 & 0.33 & 0.34 & 0.4  & 0.56 & 0.41 \\
    $\ket{T}$ & 0.97 & 0.49 & 0.49 & 0.97 & 0.97 & 0.98 \\
    \end{tabular}
\end{table}
Enforcing Hund's rule on the first site with the standard representation
    $\ket{C}$ reduces the number of appreciable CI coefficients from 170 to 50 and
    is accompanied by a significant increase in sparsity as quantified by the $L_1$
    criterion.
The increase in IPR is much smaller, because strong mixing within this reduced
    space still is necessary to resolve Hund's rule on sites $B$, $C$ and $D$.
Proceeding to $\ket{L}$ and $\ket{T}$, only six non-vanishing CSFs remain and
    their squared coefficients are shown in Table \ref{tab:psi-proj}.
\begin{table}[H]
    \caption{Weights of wave functions for the first six $S_{tot} = 8$ roots of the
        CAS(20,20) projected onto the Heisenberg manifold in the local
        sequential and balanced binary tree coupling schemes. Zeros were
        omitted to enhance legibility. Since roots two and three and four and
        five are (nearly) degenerate, mixing within the relevant subspaces is
    possible.}
    \label{tab:psi-proj}
    \begin{tabular}{@{}r cccccc@{}}
    $S_{\text{tot}} = 8$    & $\ket{\bar{\psi}_1}$ & $\ket{\bar{\psi}_2}$ & $\ket{\bar{\psi}_3}$ & $\ket{\bar{\psi}_4}$ & $\ket{\bar{\psi}_5}$ & $\ket{\bar{\psi}_6}$  \\
    \cmidrule{1-7}
        $\Delta E$ / $\SI{}{\per\cm}$ & 0 & 14 & 16 & 76 & 76 & 157 \\
    \cmidrule{1-7}
    $\ket{L} = \ket{S_{AB}, S_{ABC}}$ \\
    $\ket{3, \frac{11}{2}}$ &      & 0.48 & 0.50 &      &      &      \\[2pt]
    $\ket{4, \frac{11}{2}}$ & 0.30 &      &      &      & 0.68 &      \\[2pt]
    $\ket{4, \frac{13}{2}}$ & 0.68 &      &      &      & 0.30 &      \\[2pt]
    $\ket{5, \frac{11}{2}}$ &      & 0.07 & 0.07 & 0.44 &      & 0.41 \\[2pt]
    $\ket{5, \frac{13}{2}}$ &      & 0.21 & 0.20 & 0.10 &      & 0.48 \\[2pt]
    $\ket{5, \frac{15}{2}}$ &      & 0.23 & 0.22 & 0.45 &      & 0.09 \\

    $\ket{T} = \ket{S_{AB}, S_{CD}}$ \\
    $\ket{3, 5}$            &      & 0.48 & 0.50 &      &      &      \\
    $\ket{4, 4}$            & 0.99 &      &      &      &      &      \\
    $\ket{4, 5}$            &      &      &      &      & 0.99 &      \\
    $\ket{5, 3}$            &      & 0.50 & 0.48 &      &      &      \\
    $\ket{5, 4}$            &      &      &      & 0.99 &      &      \\
    $\ket{5, 5}$            &      &      &      &      &      & 0.99 \\
    \end{tabular}
\end{table}
Although sparsity is maximised in $\ket{T}$, as would be expected from equation
    \eqref{eq:4j-heisenberg}, the largest gain is already realised by
    recoupling from $\ket{C}$ to $\ket{L}$, where Hund's rule can be expressed on every site.
This example makes a case in point for the $\ket{L}$ basis, because the
    redox cycles in nature mostly feature metal centres of mixed oxidation
    states for which $\ket{T}$ CSFs are not expected to be more descriptive.
Nevertheless, we presume that non-sequential couplings could be advantageous for
    partitioning large magnetic systems into smaller units whenever the subsystem
    exhibits a particular symmetry, e.g. treating the FeMo cofactor in terms of
    two magnetically coupled cuboids.
Note that while equation (16) does not hold for inequivalent magnetic sites,
    the recoupling of CI vectors into local spin–adapted representations remains
    applicable.
In these cases, a new recoupling diagram, similar to the example
    provided in section 2, can be drawn and reduced to a sequence of Racah symbols
    and Kronecker deltas.

\section{Conclusion}
In this letter, we extended the effective Hamiltonian formalism
    to polynuclear transition metal complexes with local spins
    greater than one half by means of recoupling transformations.
The method was used to assess a spin model for a \ce{[CaMn^{(IV)}3O4]} cubane
    with minimal and extended active space calculations.
Enlarging the active space was shown to qualitatively change the magnetism from a
    three parameter to a two parameter interaction, illustrating the utility and
    importance of an exact mapping between ab initio and model wave functions.
An extension to sparse solvers like DMRG or FCIQMC is readily possible, if CI
    coefficients can be efficiently extracted.
We also discussed how chains of local spin-adapted functions can be transformed
    into non-sequential, tree-like orderings to express additional
    permutational symmetries of model Hamiltonians in ab initio wave functions.
Overall, the increased compactness of CI solutions in nonstandard
    coupling schemes portends a promising future for their use in sparse solvers
    like FCIQMC or as reference states for perturbation theory.
It remains to be seen whether the additional sparsity compared to the
    standard basis outweighs the overhead of more complicated Hamiltonian matrix
    element evaluation.\cite{gould1986a,gould1986b}

\section*{Appendix A. Recapitulation of Angular Momentum Diagrams}
Clebsch-Gordan coefficients, $\braket{j_1 m_1 ; j_2 m_2|j_3 m_3}$, are the
    building blocks of the Racah-Wigner calculus.
The angular momenta fulfil the triangle condition:
\begin{equation}
    \begin{split}
        |j_1 - j_2| \leq j_3 \leq j_1 + j_2
    \end{split}
\end{equation}
    and sum to an integer $j_1 + j_2 + j_3 \in \mathbb{Z}^{0\leq}$.
In graphical notation, they are represented by three lines connected to a
    semicircle vertex.
Standard and contrastandard states are denoted by single and double arrows,
    respectively.
SU(2) invariants are formed by summing over all magnetic quantum numbers of a
    standard/contrastandard product.
Since bra/ket and ket/bra contractions are invariant under simultaneous
    time-reversal, these contracted lines are not oriented, see also
    equations (54) and (55) in reference \citenum{wormer2006}.
Note that coupling more than two angular momenta implies a summation over the
    magnetic quantum numbers of intermediate $j$, e.g.:
    \begin{equation}
        \braket{j_1 m_1 ; j_2 m_2 ; j_3 m_3|((j_1 j_2) j_3) j m} =
        \sum_{m_{12}} \braket{j_1 m_1 ; j_2 m_2|j_{12} m_{12}}
                          \braket{j_{12} m_{12} ; j_3 m_3|j m}.
    \end{equation}
Whereas recoupling problems are conveniently expressed in terms of generalised
    Clebsch-Gordan coefficients, owing to their cumbersome symmetries, it is
    convenient for the diagrammatic approach to use more symmetric 3-$jm$ symbols
    instead.
Following chapter 3.6 of reference \citenum{wormer2006}, this conversion
    involves five steps (always referring to the initial diagram):\cite{wormer2006}
\begin{enumerate}[(i)]
    \item change the sign of the vertex,
    \item keeping their direction, exchange all double arrows with single arrows,
    \item add a factor $\sqrt{2j + 1}$ for the unique line, i.e. the line
          of bra or ket type appearing only once,
    \item add a factor $(-1)^{2j}$ for the first line encountered in going
          from the unique single (ket)/double (bra) arrow line into the direction indicated
          by the vertex sign (for a ket: $-$ clockwise / $+$ counter clockwise,
	  for a bra: $-$ counter clockwise / $+$ clockwise),
    \item add a factor $(-1)^{2j}$ for every time-reversed bra (outgoing double arrow).
\end{enumerate}
For example:
\begin{center}
    \begin{tikzpicture}
        \node at (-0.5,0.3) {$j_3$};
        \node at (0.5,0.6) {$j_1$};
        \node at (0.5,-0.6) {$j_2$};
        \node at (0,0) [cgvertex, shape border rotate = -90, draw, label=right:$-$] {};
        % polar coordinates
        \draw [doublearrow] (180:0.05) -- (180:1);
        \draw [singlearrow] (30:0.05)  -- (30:1) ;
        \draw [singlearrow] (330:0.05) -- (330:1);
    \end{tikzpicture}
    \begin{tikzpicture}
        \node at (-0.5,0.3) {$j_3$};
        \node at (0.5,0.6) {$j_1$};
        \node at (0.5,-0.6) {$j_2$};
        \node at (-1.3,0) {$=$};
        \node at (0,0) [vertex, draw, label=below:$+$] {};
        \draw [singlearrow] (180:0.05) -- (180:1);
        \draw [singlearrow] (30:0.05)  -- (30:1)    ;
        \draw [singlearrow] (330:0.05) -- (330:1)   ;
    \end{tikzpicture}
\end{center}
\begin{equation}
    \braket{\theta(j_3m_3) |j_1m_1 \text{; } j_2m_2} = (-1)^{2j_2 + 2j_3} \sqrt{2j_3 + 1}
    \begin{pmatrix} j_1 & j_3 & j_2 \\ m_1 & m_3 & m_2 \end{pmatrix}
\end{equation}
Changing the node sign on a 3-$jm$ symbol, corresponding to an odd permutation
    of columns, incurs a factor of $(-1)^{j_1 + j_2 + j_3}$.
Unlike the bra/ket and ket/bra contractions of Clebsch-Gordan coefficients,
    the lines of contracted 3-$jm$ symbols are oriented.
Arrows on contracted lines can be inverted, introducing a factor of $(-1)^{2j}$.
A diagram with two external lines can be closed as follows:
\begin{center}
    \begin{tikzpicture}
        \draw (0,0) -- (1,0) -- (1,2.5) -- (0,2.5) -- cycle;
        \draw [singlearrow] (2,0.625) -- (1,0.625);
        \draw [singlearrow] (1,1.875) -- (2,1.875);
        \node at (1.5,2.25) {$j_1 m_1$};
        \node at (1.5,1.0) {$j_2 m_2$};
    \end{tikzpicture}
    \begin{tikzpicture}
        \node (equation) at (-1.75,1.25) {$= \delta_{j_1 j_2} \delta_{m_1 m_2} \frac{1}{(2j_1 + 1)}$};
        \draw (0,0) -- (1,0) -- (1,2.5) -- (0,2.5) -- cycle;
        \node (left_lower) at (0.87,0.625) {};
        \node (left_upper) at (0.87,1.875) {};
        \draw [singlearrow] (left_upper) to [bend left=90] (left_lower);
        \node at (1.5,1.9) {$j_1$};
    \end{tikzpicture}
\end{center}
From a contracted graph, a sequence of line separations leads to irreducible
    3$n$-$j$ symbols.
In the main text, we use the two and three line separation theorems:
\begin{equation}
    \begin{tikzpicture}
        \draw (0,0) -- (1,0) -- (1,2.5) -- (0,2.5) -- cycle;
        \draw [singlearrow] (1,0.625) -- (2,0.625);
        \draw [singlearrow] (1,1.875) -- (2,1.875);
        \draw (3,0) -- (2,0) -- (2,2.5) -- (3,2.5);
        \node at (1.5,2.25) {$j_1$};
        \node at (1.5,1.0) {$j_2$};
    \end{tikzpicture}
    \begin{tikzpicture}
        \node (equation) at (-1.5,1.25) {$= \delta_{j_1 j_2}\frac{1}{(2j_1 + 1)}$};
        \draw (0,0) -- (1,0) -- (1,2.5) -- (0,2.5) -- cycle;
        \draw (3.5,0) -- (2.5,0) -- (2.5,2.5) -- (3.5,2.5);
        \node (left_lower) at (0.87,0.625) {};
        \node (left_upper) at (0.87,1.875) {};
        \node (right_lower) at (2.62,0.625) {};
        \node (right_upper) at (2.62,1.875) {};
        \draw [singlearrow] (left_lower) to [bend right=90] (left_upper);
        \draw [singlearrow] (right_lower) to [bend left=90] (right_upper);
        \node at (1.5,1.9) {$j_1$};
        \node at (2.1,1.9) {$j_1$};
    \end{tikzpicture}
\end{equation}

\begin{equation}
    \begin{tikzpicture}
        \node at (1.5, 1.55) {$j_1$};
        \node at (1.5, 0.95) {$j_2$};
        \node at (1.5, 0.35) {$j_3$};
        \draw (0,0) -- (1,0) -- (1,2.5) -- (0,2.5) -- cycle;
        \draw [singlearrow] (1,0.625) -- (2,0.625);
        \draw [singlearrow] (1,1.250) -- (2,1.250);
        \draw [singlearrow] (1,1.875) -- (2,1.875);
        \draw (3,0) -- (2,0) -- (2,2.5) -- (3,2.5);
    \end{tikzpicture}
    \begin{tikzpicture}
        \node (equation) at (-1.25,1.25) {$=$};
        \draw (0,0) -- (1,0) -- (1,2.5) -- (0,2.5) -- cycle;
        \node (positive) at (2.25,1.250)    [vertex, label=above:$+$] {};
        \node (negative) at (2.75,1.250)    [vertex, label=above:$-$] {};
        \draw [singlearrow] (1,0.625) -- (positive);
        \draw [singlearrow] (1,1.250) -- (positive);
        \draw [singlearrow] (1,1.875) -- (positive);
        \draw [singlearrow] (negative) -- (4,0.625);
        \draw [singlearrow] (negative) -- (4,1.250);
        \draw [singlearrow] (negative) -- (4,1.875);
        \draw (5,0) -- (4,0) -- (4,2.5) -- (5,2.5);
        \node at (1.5, 1.95) {$j_1$};
        \node at (1.2, 1.05) {$j_2$};
        \node at (1.5, 0.55) {$j_3$};
        \node at (3.5, 1.95) {$j_1$};
        \node at (3.8, 1.05) {$j_2$};
        \node at (3.5, 0.55) {$j_3$};
    \end{tikzpicture}
\end{equation}
Closed and open boxes represent diagrams with fully and non-contracted lines,
    respectively.

The final expressions are written in terms of $3j$ and $6j$ symbols,
    the former of which is defined as:
\begin{center}
    \begin{tikzpicture}[scale=1.3]
        % vertices
        \node at (0,0)    [vertex, label=left:$-$] {};
        \node at (2,0)    [vertex, label=right:$+$] {};
        % lines
        \draw [singlearrow] (0,0) -- (2,0);
        \draw [singlearrow] (0,0) to[out= 70, in= 110] (2,0);
        \draw [singlearrow] (0,0) to[out=-70, in=-110] (2,0);
        %labels
        \node at (1.0, 0.75) {$j_3$};
        \node at (1.0, 0.20) {$j_2$};
        \node at (1.0,-0.35) {$j_1$};
    \end{tikzpicture}
\end{center}
\begin{equation}
    \equiv \begin{Bmatrix}
        j_1 & j_2 & j_3
    \end{Bmatrix} =
    \begin{cases}
        1 , & \text{if } |j_1 - j_2| \leq j_3 \leq j_1 + j_2 \\
        0 , & \text{else}.
    \end{cases}
\end{equation}
The $6j$ symbol is a contraction of four 3-$jm$ symbols:
\begin{center}
    \begin{tikzpicture}[scale=1.3]
    ]
        % 6j frame
        \draw [singlearrow] (0,0) -- (2,0);
        \draw [singlearrow] (2,0) -- (1,2);
        \draw [singlearrow] (1,2) -- (0,0);
        \draw [singlearrow] (1,0.75) -- (0,0);
        \draw [singlearrow] (1,0.75) -- (2,0);
        \draw [singlearrow] (1,0.75) -- (1,2);
        % vertices
        \node at (0,0)    [vertex, label=below:$+$] {};
        \node at (2,0)    [vertex, label=below:$+$] {};
        \node at (1,2)    [vertex, label=above:$+$] {};
        \node at (1,0.75) [vertex, label=right:$+$] {};
        % labels
        \node at (1.15,1.15) [] {$j_1$};
        \node at (1.3,0.35) [] {$j_3$};
        \node at (0.7,0.35) [] {$j_2$};
        \node at (1,-0.25) [] {$j_4$};
        \node at (0.25,1.05) [] {$j_6$};
        \node at (1.75,1.05) [] {$j_5$};
    \end{tikzpicture}
    \begin{tikzpicture}[scale=1.3]
        % 6j frame
        \draw [singlearrow] (2,0) -- (0,0);
        \draw [singlearrow] (1,2) -- (2,0);
        \draw [singlearrow] (0,0) -- (1,2);
        \draw [singlearrow] (1,0.75) -- (0,0);
        \draw [singlearrow] (1,0.75) -- (2,0);
        \draw [singlearrow] (1,0.75) -- (1,2);
        % vertices
        \node at (0,0)    [vertex, label=below:$-$] {};
        \node at (2,0)    [vertex, label=below:$-$] {};
        \node at (1,2)    [vertex, label=above:$-$] {};
        \node at (1,0.75) [vertex, label=right:$-$] {};
        % labels
        \node at (1.15,1.15) [] {$j_1$};
        \node at (1.3,0.35) [] {$j_2$};
        \node at (0.7,0.35) [] {$j_3$};
        \node at (1,-0.25) [] {$j_4$};
        \node at (0.25,1.05) [] {$j_5$};
        \node at (1.75,1.05) [] {$j_6$};
    \end{tikzpicture}
\end{center}
\begin{equation}
    \equiv \begin{Bmatrix}
        j_1 & j_2 & j_3 \\
        j_4 & j_5 & j_6
    \end{Bmatrix}
\end{equation}
In standard form, all vertices have the same sign.
The inner lines either all point towards or away from the centre, whereas the
    outer arrows follow the direction indicated by the vertex sign.
Inverting all signs and outer arrow directions is equivalent to reflecting the
    symbol on an outer axis.
Arbitrary column permutations or the pairwise inversion of bottom
    and top elements leave the $6j$ symbol invariant, e.g.:
\begin{equation}
    \begin{Bmatrix}
        j_1 & j_2 & j_3 \\
        j_4 & j_5 & j_6
    \end{Bmatrix} =
    \begin{Bmatrix}
        j_6 & j_2 & j_4 \\
        j_3 & j_5 & j_1
    \end{Bmatrix}.
\end{equation}
Many programming languages provide optimised libraries or wrappers to obtain
    the numerical values of $3n$-$j$ symbols.\cite{fujiipy3nj}

%% The "Acknowledgment" section can be given in all manuscript
%% classes.  This should be given within the "acknowledgment"
\begin{acknowledgement}
Funding was provided by the Max Planck Society. N. A. Bogdanov thanks Huanchen
Zhai for discussions on how to extract CI coefficients from MPS wave functions.
Both authors thank the anonymous reviewer for bringing their attention to
fragmentation-based wave function approaches, such as the Localised Active
Space.
\end{acknowledgement}

% \begin{suppinfo}

% The Supporting Information contain in PDF format: (A) Recapitulation of
% angular momentum diagram rules (B) analytical and numerical Hamiltonians for
% the $S = 10, 9, 7, 6$ manifolds and as a compressed file (ZIP): (C)
% \texttt{OpenMolcas} CASSCF natural orbitals and CI vectors in the unitary group
% formalism with localised, site ordered orbitals.

% \end{suppinfo}

% \section*{TOC Graphic}
% \includegraphics[height=5cm]{graphical_abstract.pdf}

{ \small
\bibliography{references}
}

\end{document}

\singlespacing
%\linespread{1.05}
% \begin{tocentry}

% Some journals require a graphical entry for the Table of Contents.
% This should be laid out ``print ready'' so that the sizing of the
% Inside the \texttt{tocentry} environment, the font used is Helvetica
% 8\,pt, as required by \emph{Journal of the American Chemical
% The surrounding frame is 9\,cm by 3.5\,cm, which is the maximum
% permitted for  \emph{Journal of the American Chemical Society}
% graphical table of content entries. The box will not resize if the
% content is too big: instead it will overflow the edge of the box.
%
% This box and the associated title will always be printed on a
% separate page at the end of the document.
%
% \end{tocentry}

% NOTE: The abstract is commented out for a reason.
%       Please leave it like this!
% \begin{abstract}
% % Abstract can only have 75 words! Keep it in mind when applying changes.
% \end{abstract}
\tableofcontents

\nopagebreak
\section{RASSCF orbitals for the \ce{[CaMn3^{(IV)}O4]} Cubane}
\begin{figure}[H]
    \includegraphics[width=0.245\textwidth]{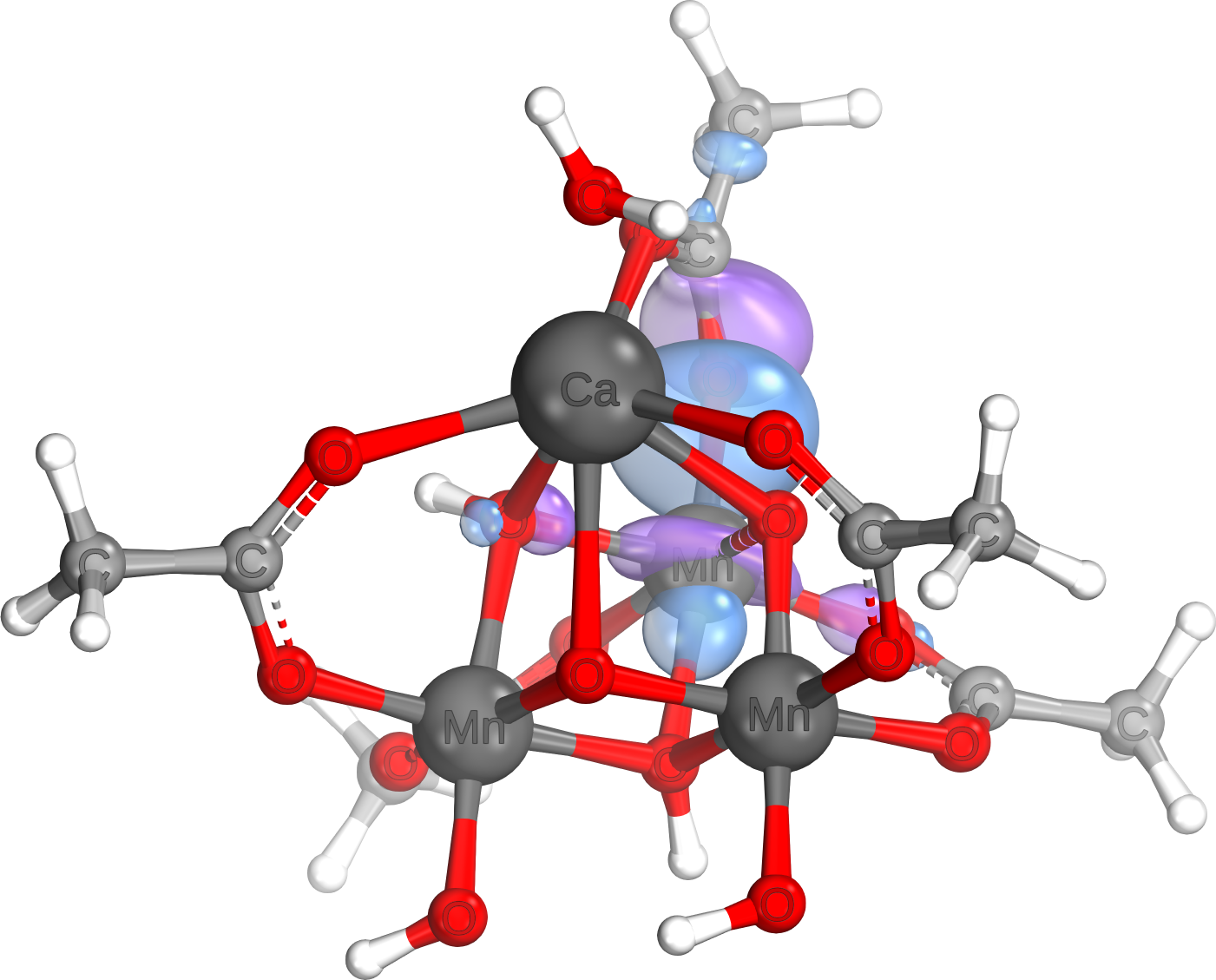}
    \includegraphics[width=0.245\textwidth]{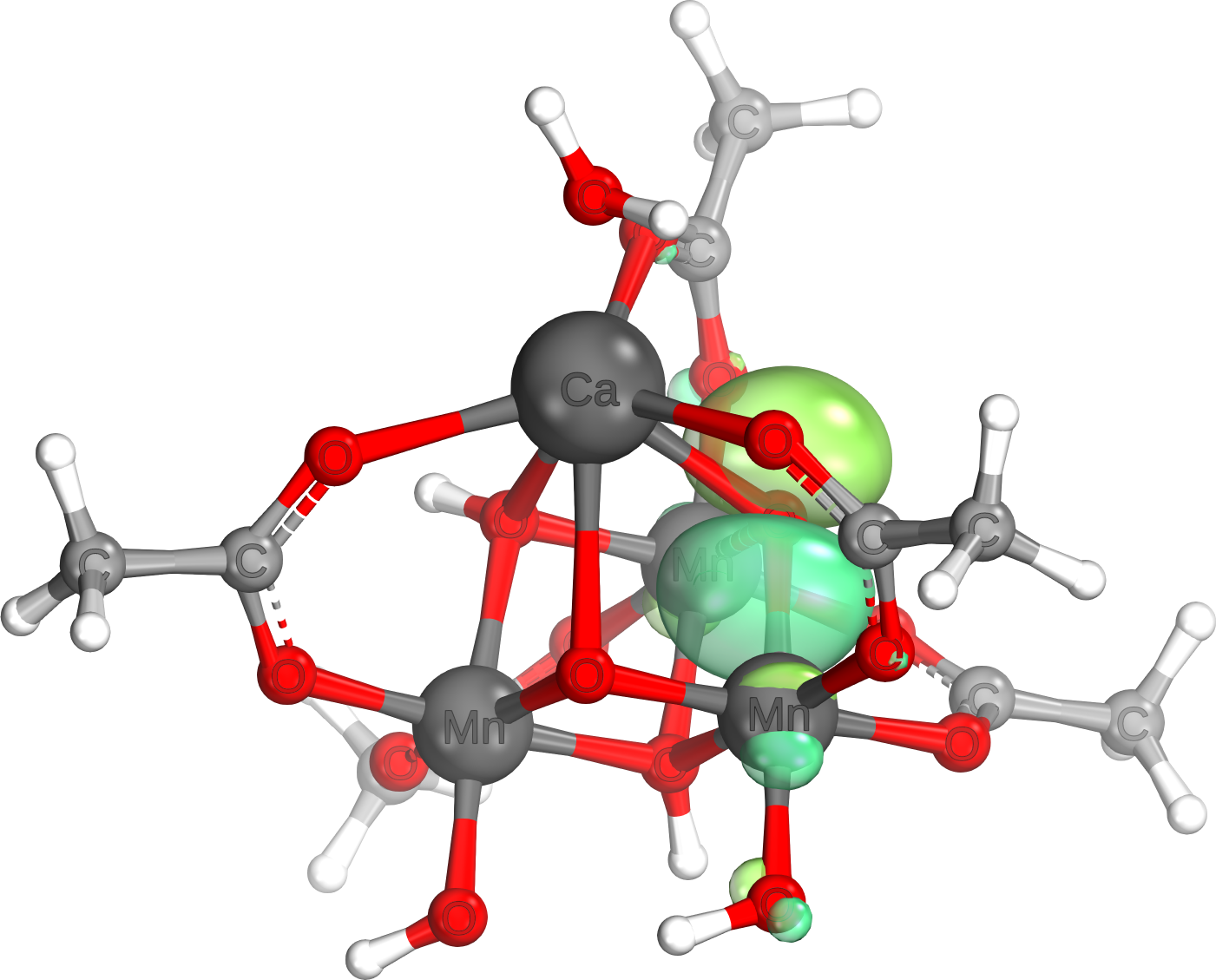}
    \includegraphics[width=0.245\textwidth]{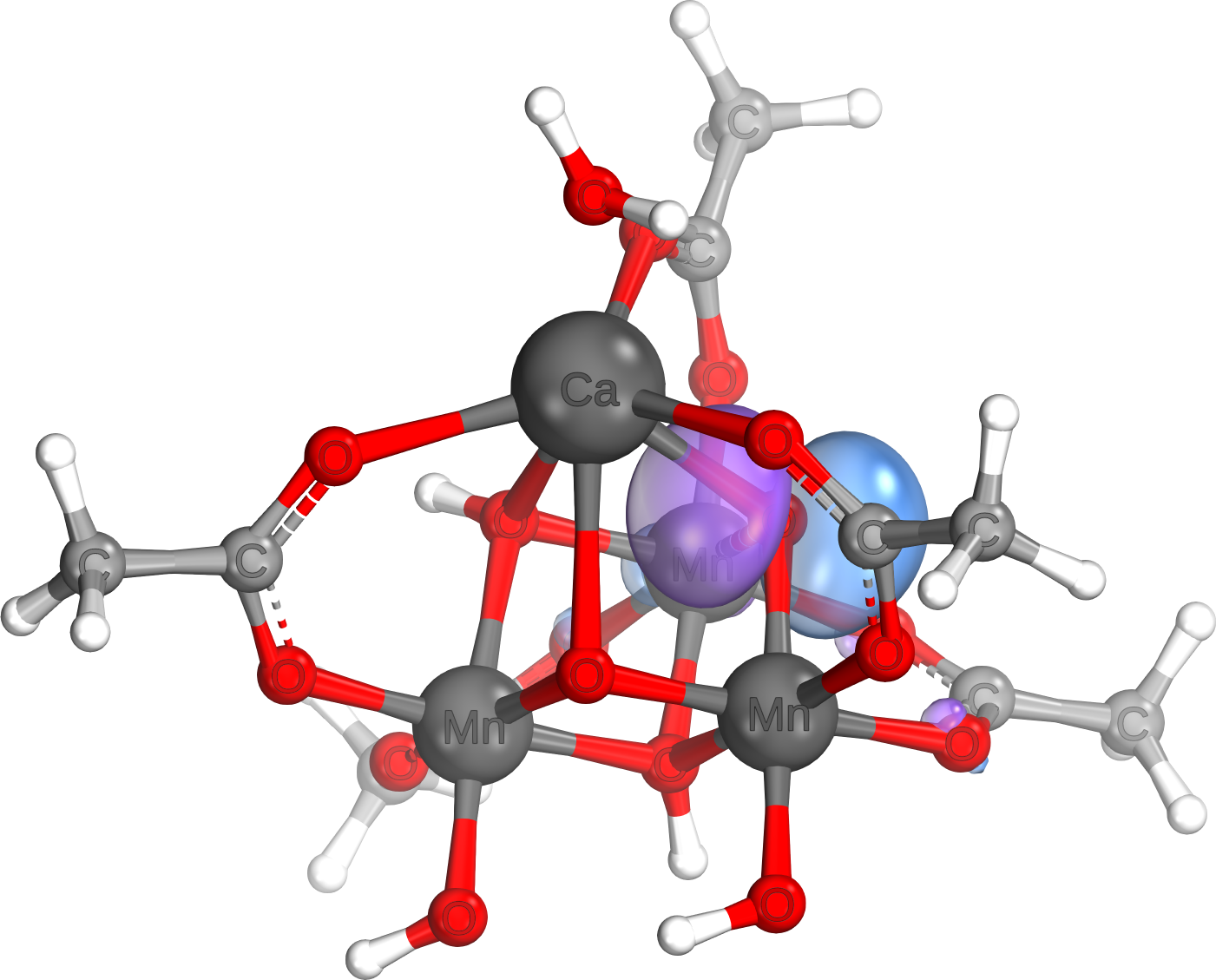}
    \includegraphics[width=0.245\textwidth]{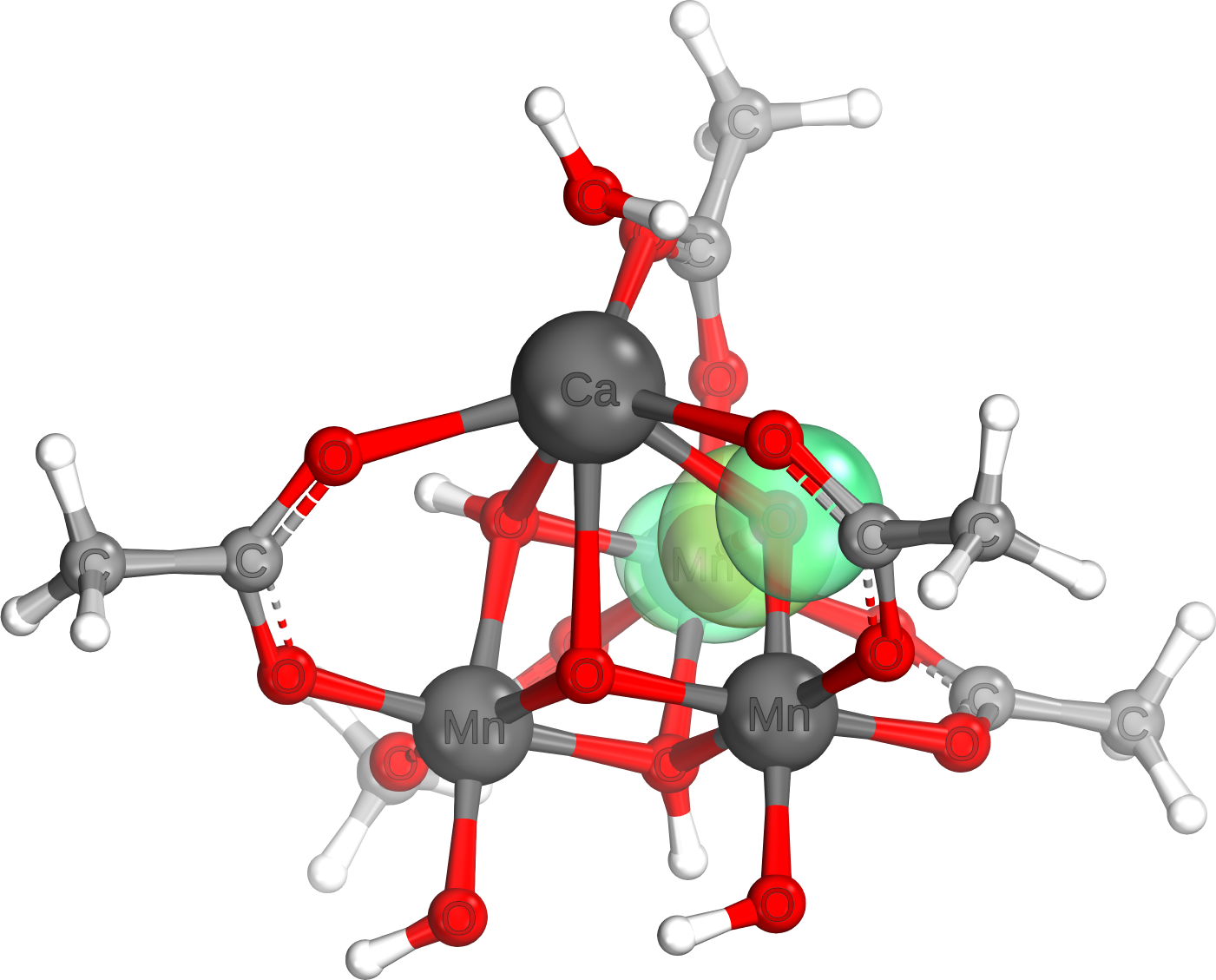} \\
    \includegraphics[width=0.245\textwidth]{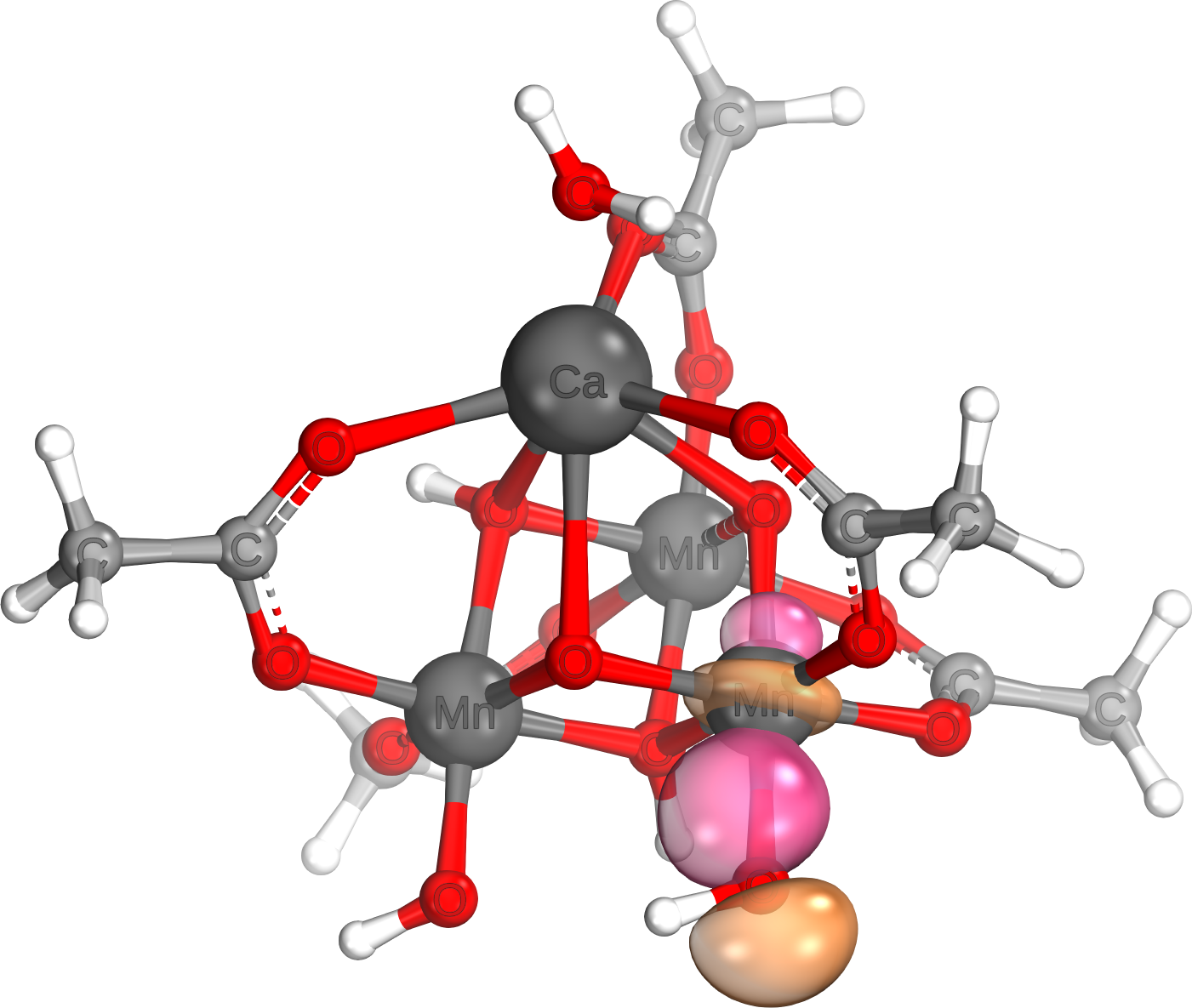}
    \includegraphics[width=0.245\textwidth]{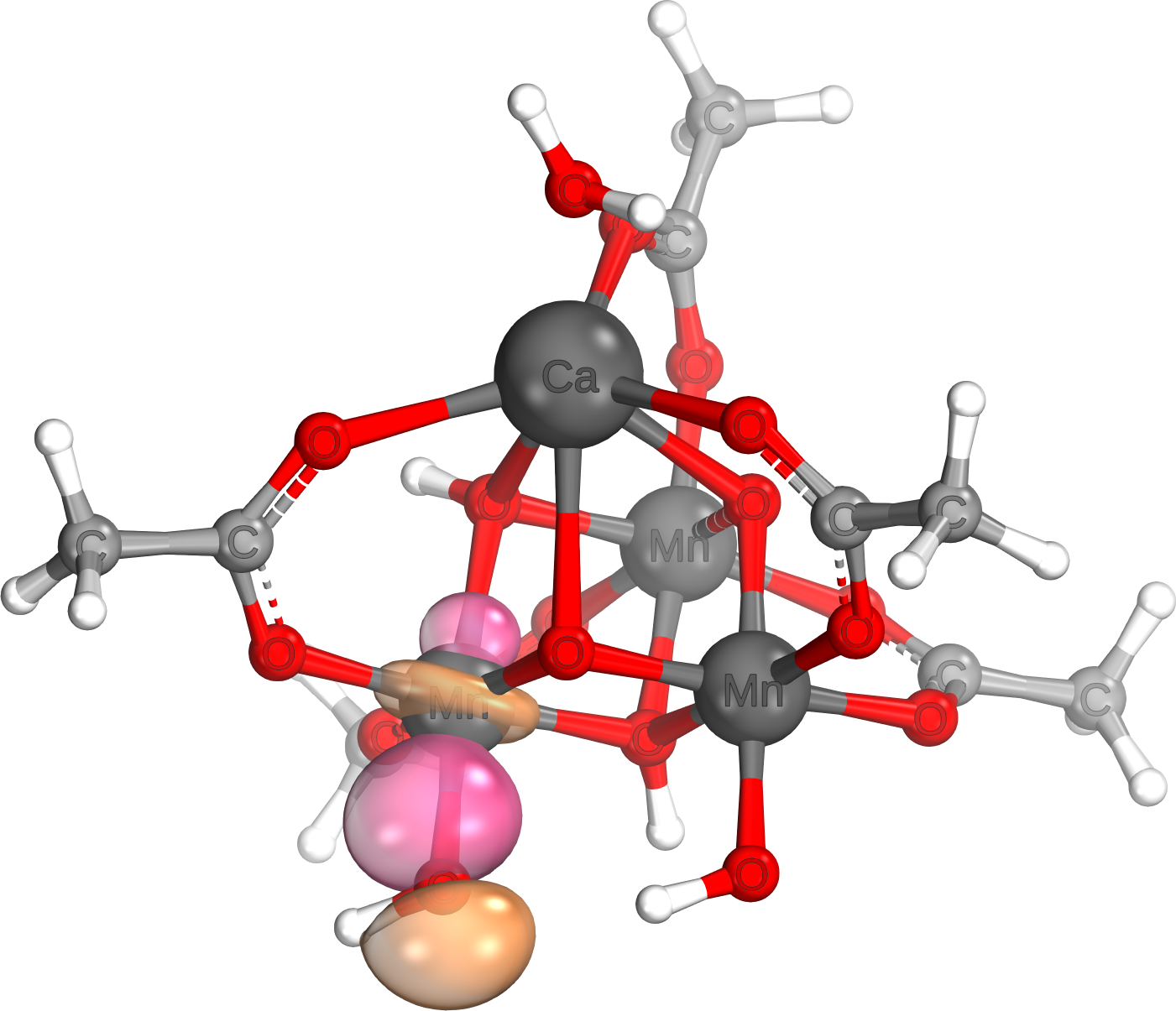}
    \includegraphics[width=0.245\textwidth]{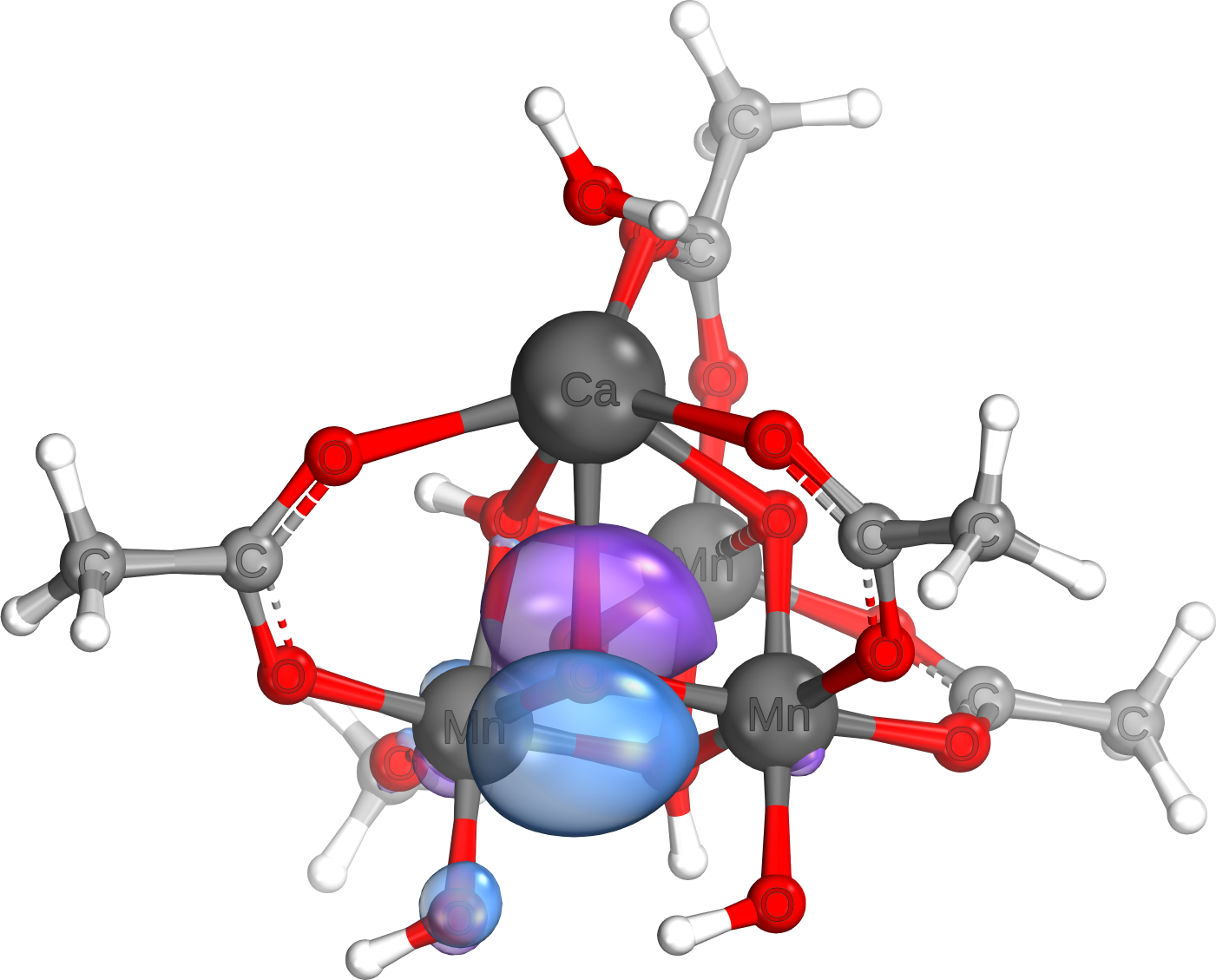}
    \includegraphics[width=0.245\textwidth]{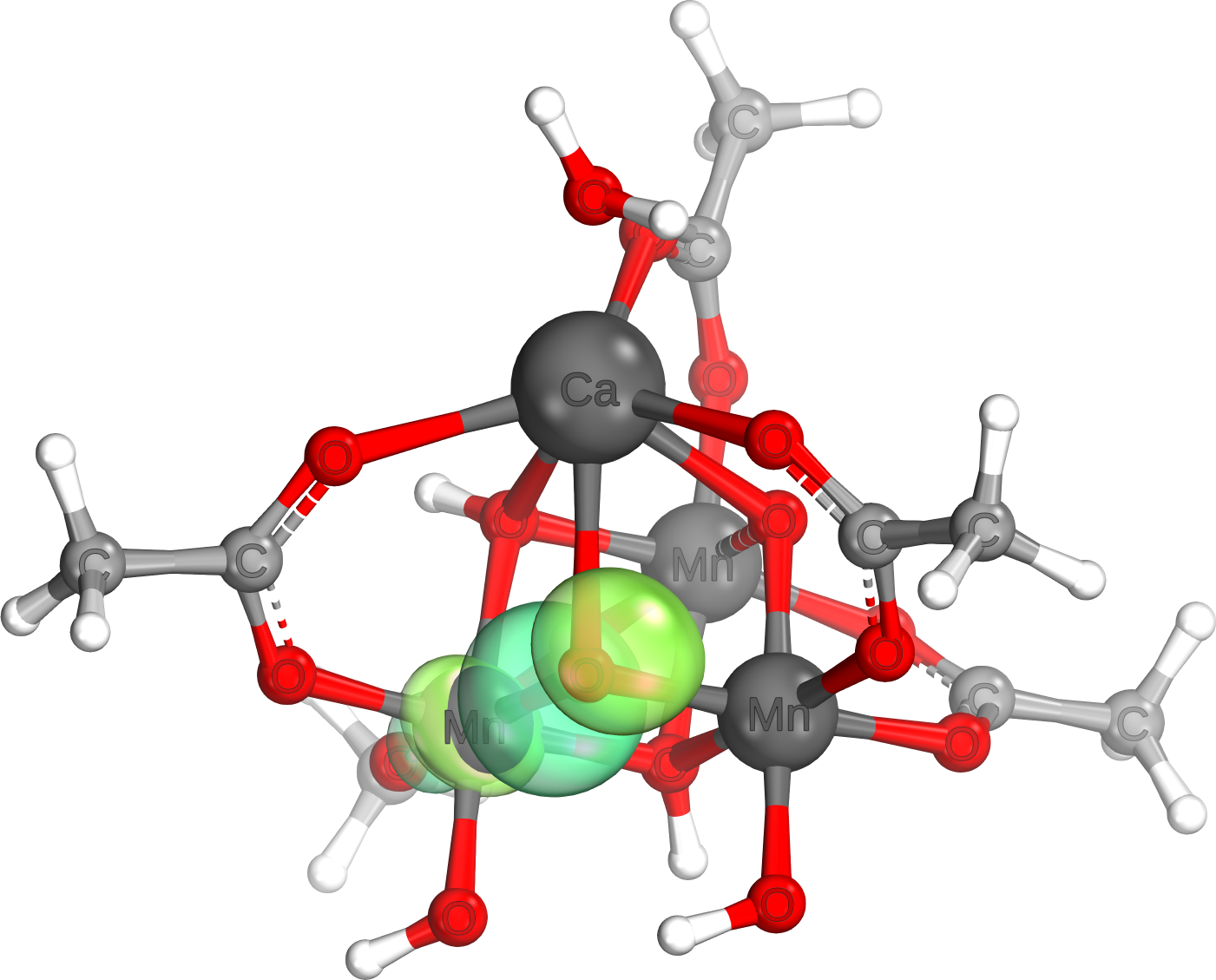} \\
    \includegraphics[width=0.245\textwidth]{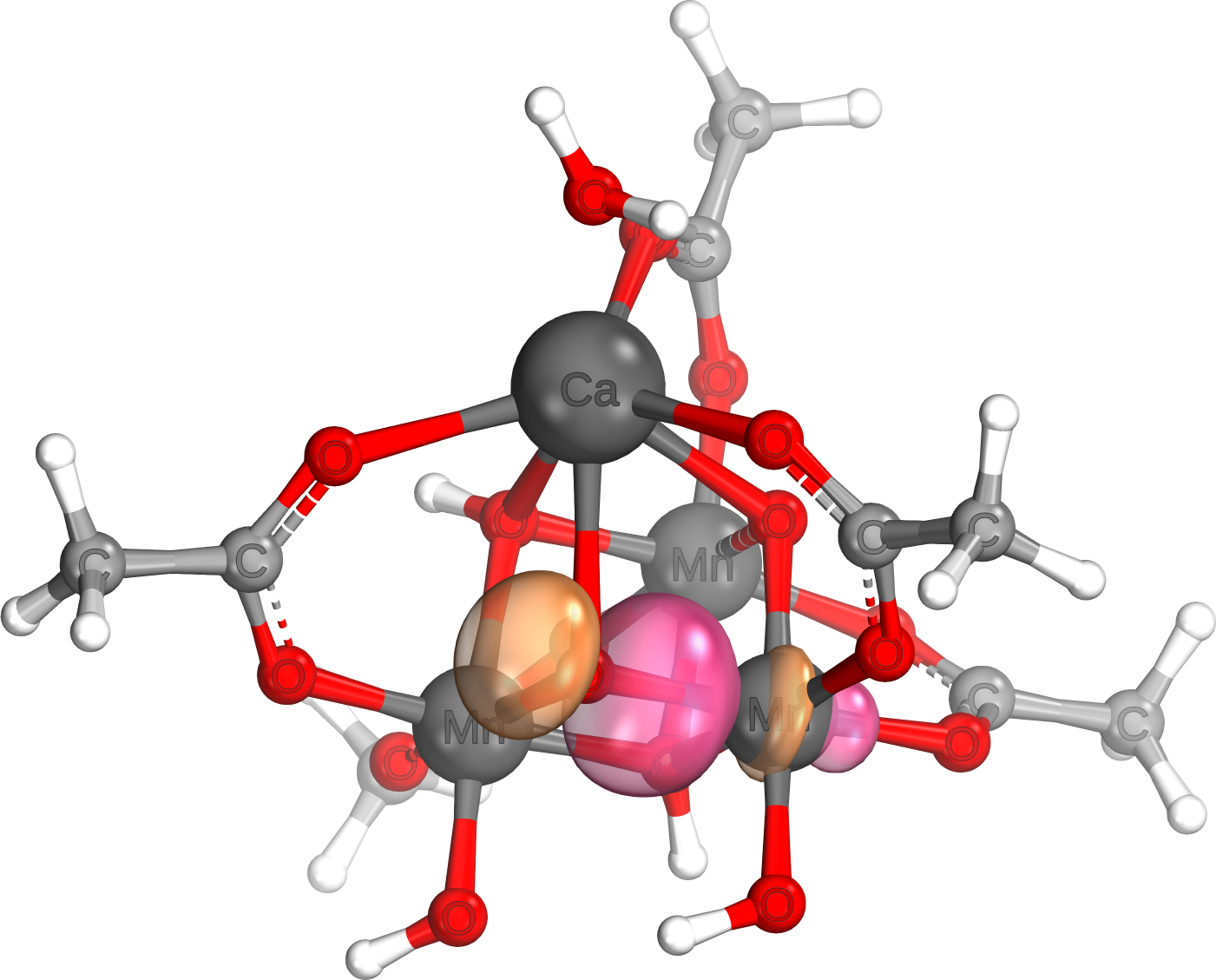}
    \includegraphics[width=0.245\textwidth]{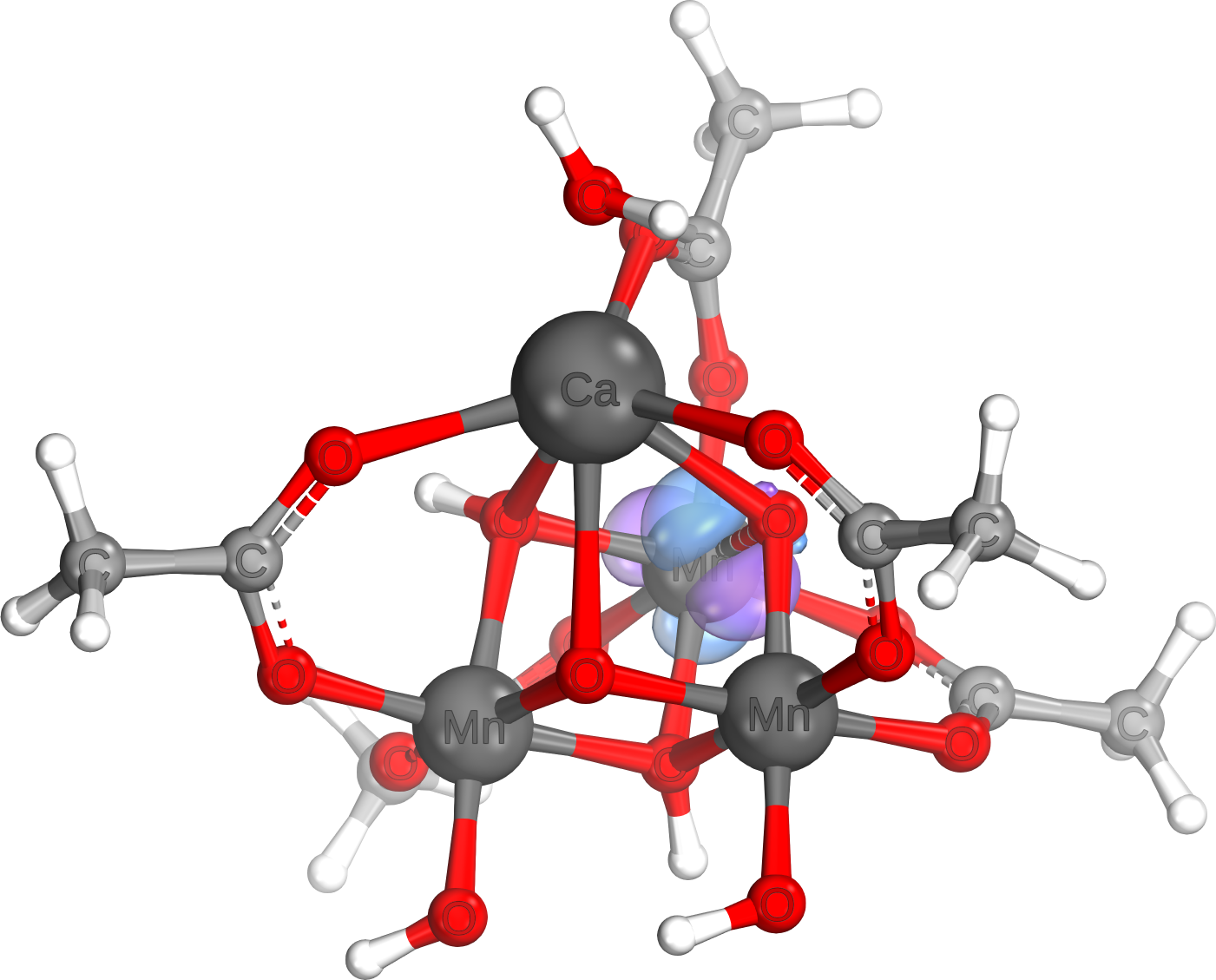}
    \includegraphics[width=0.245\textwidth]{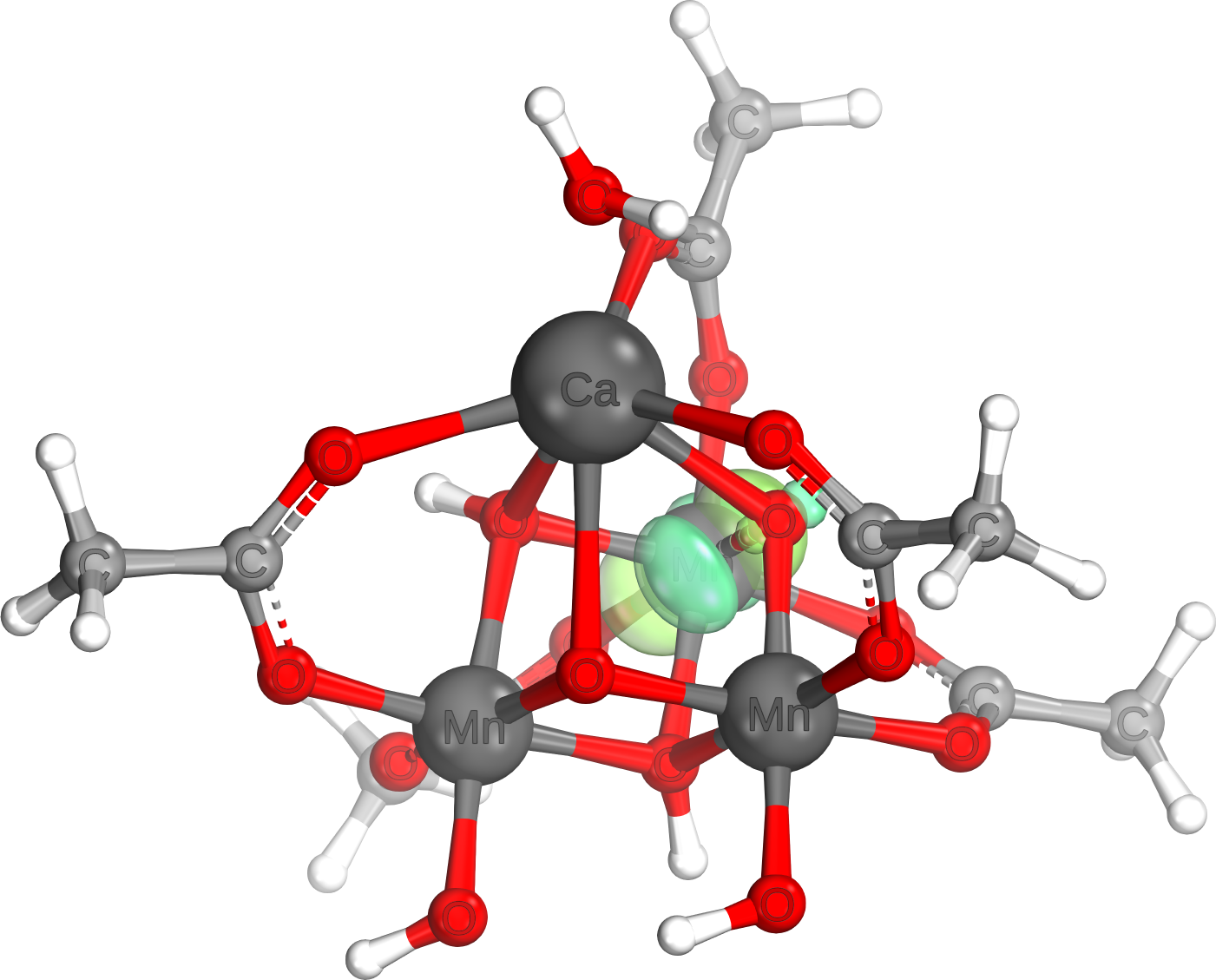}
    \includegraphics[width=0.245\textwidth]{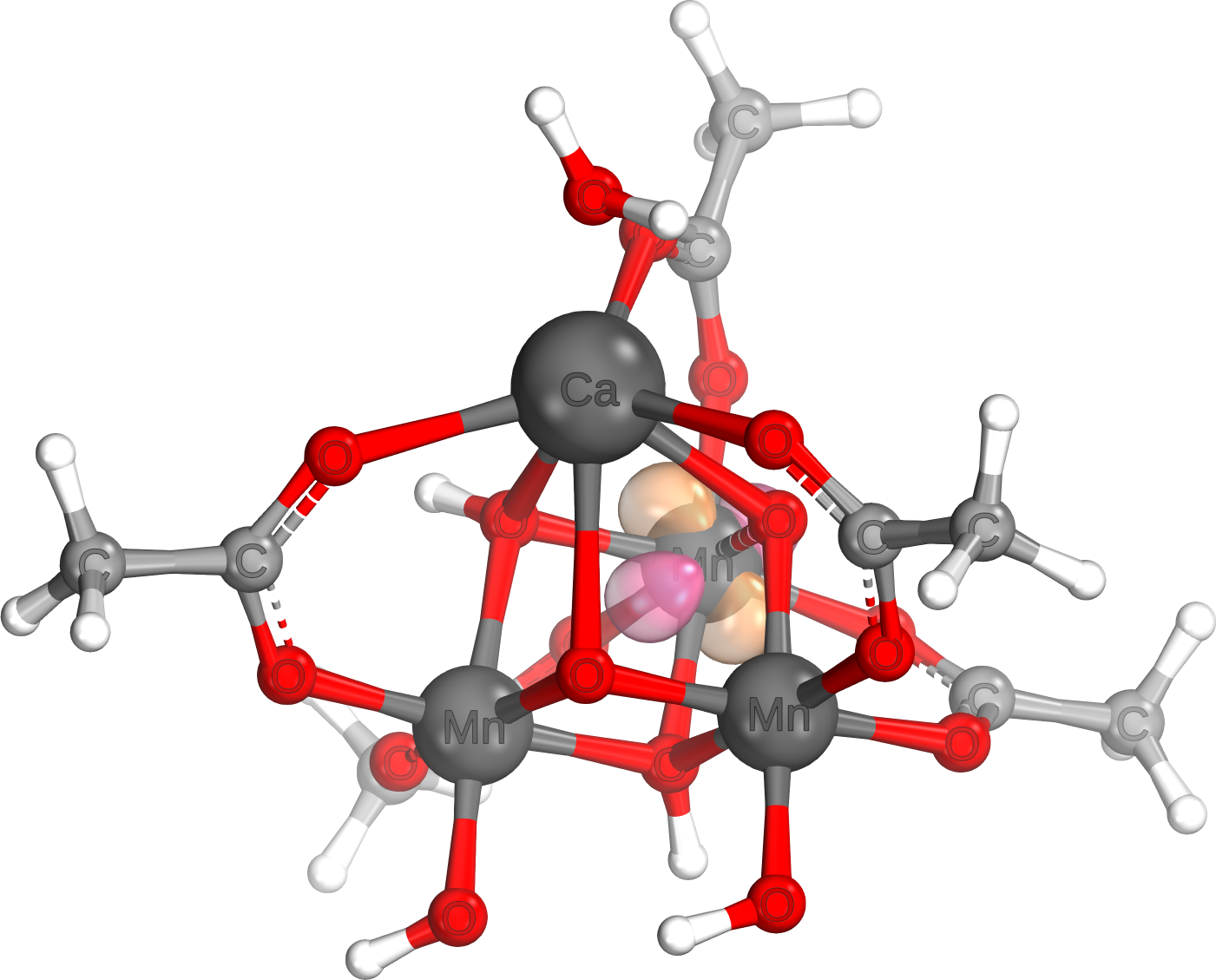} \\
    \includegraphics[width=0.245\textwidth]{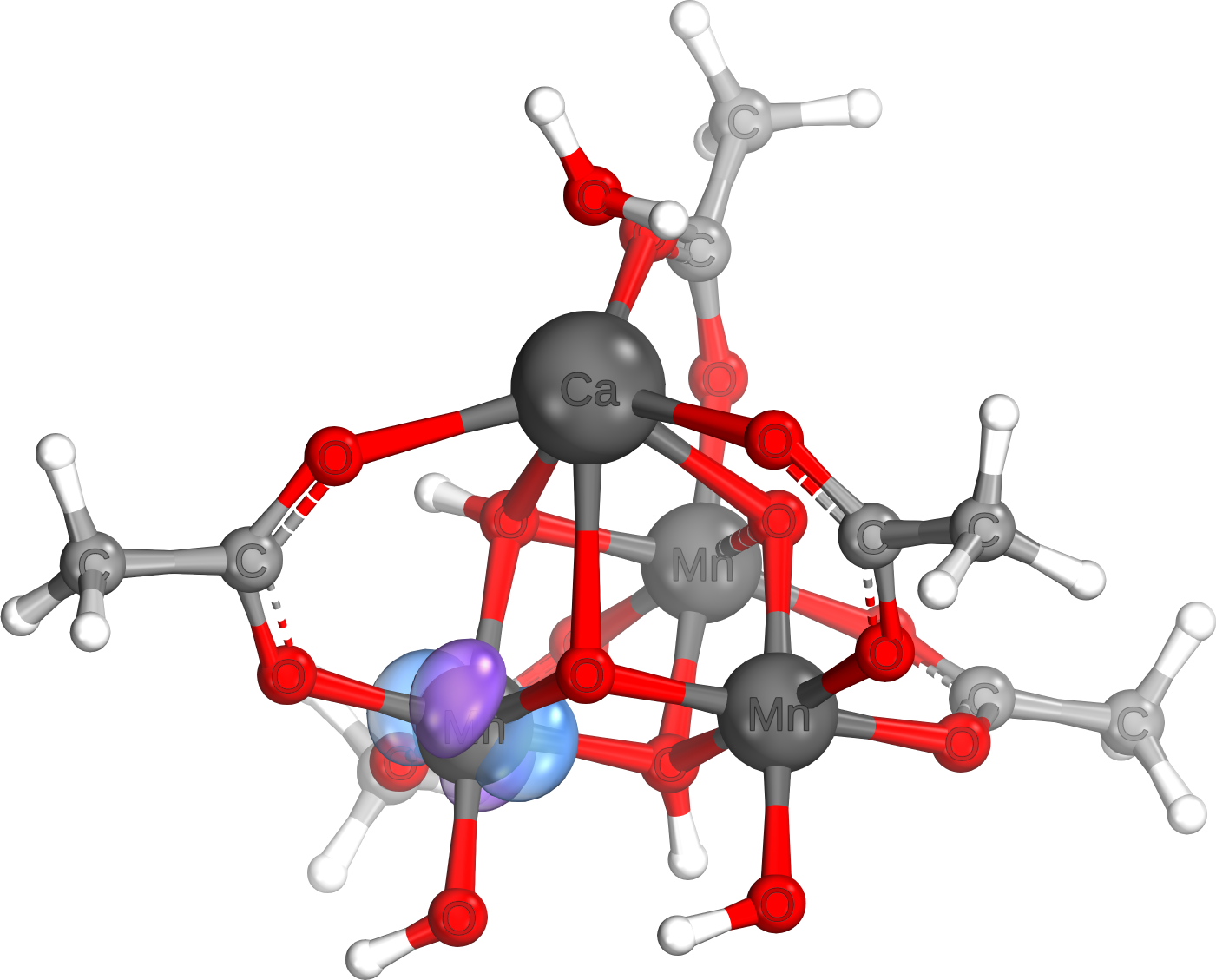}
    \includegraphics[width=0.245\textwidth]{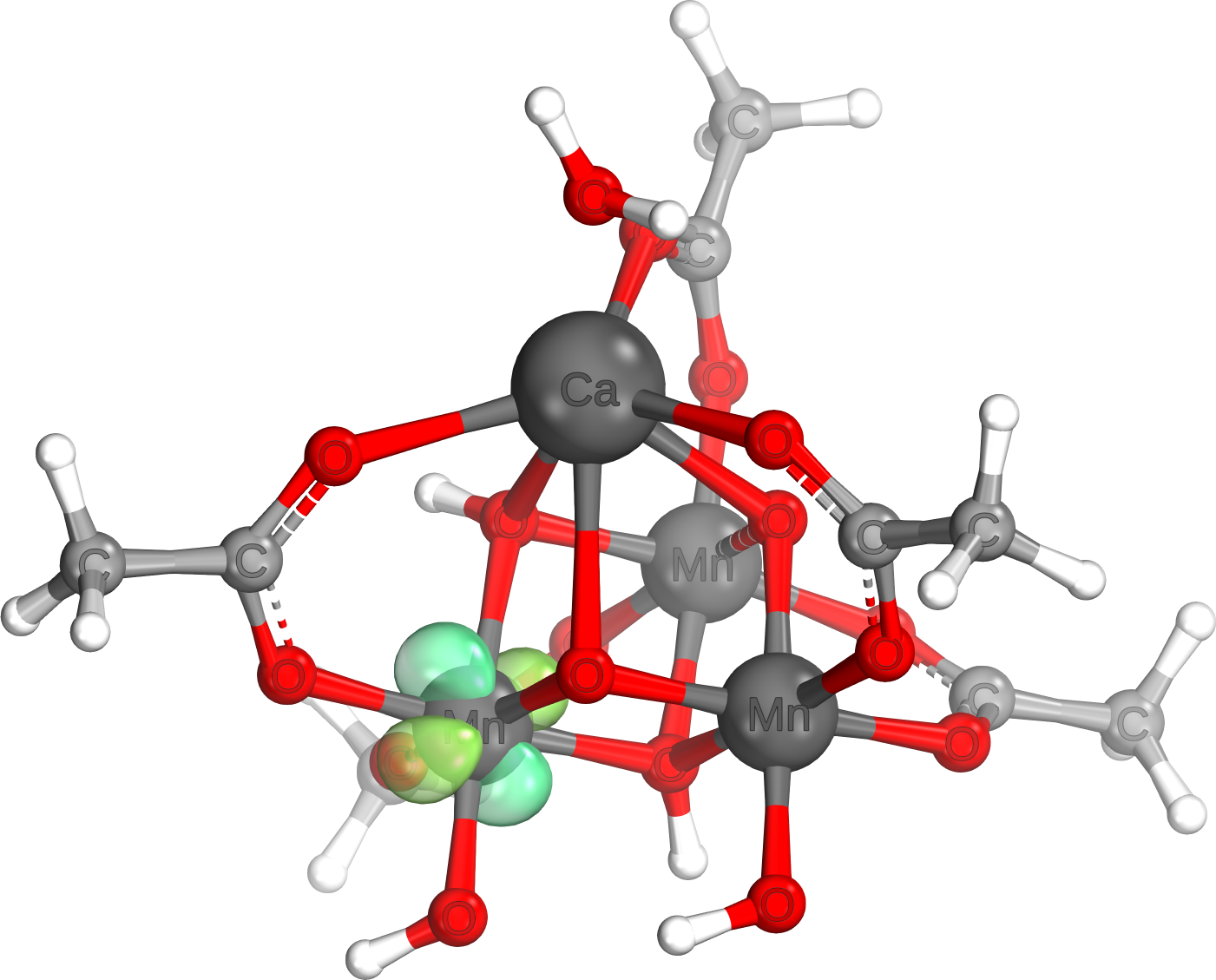}
    \includegraphics[width=0.245\textwidth]{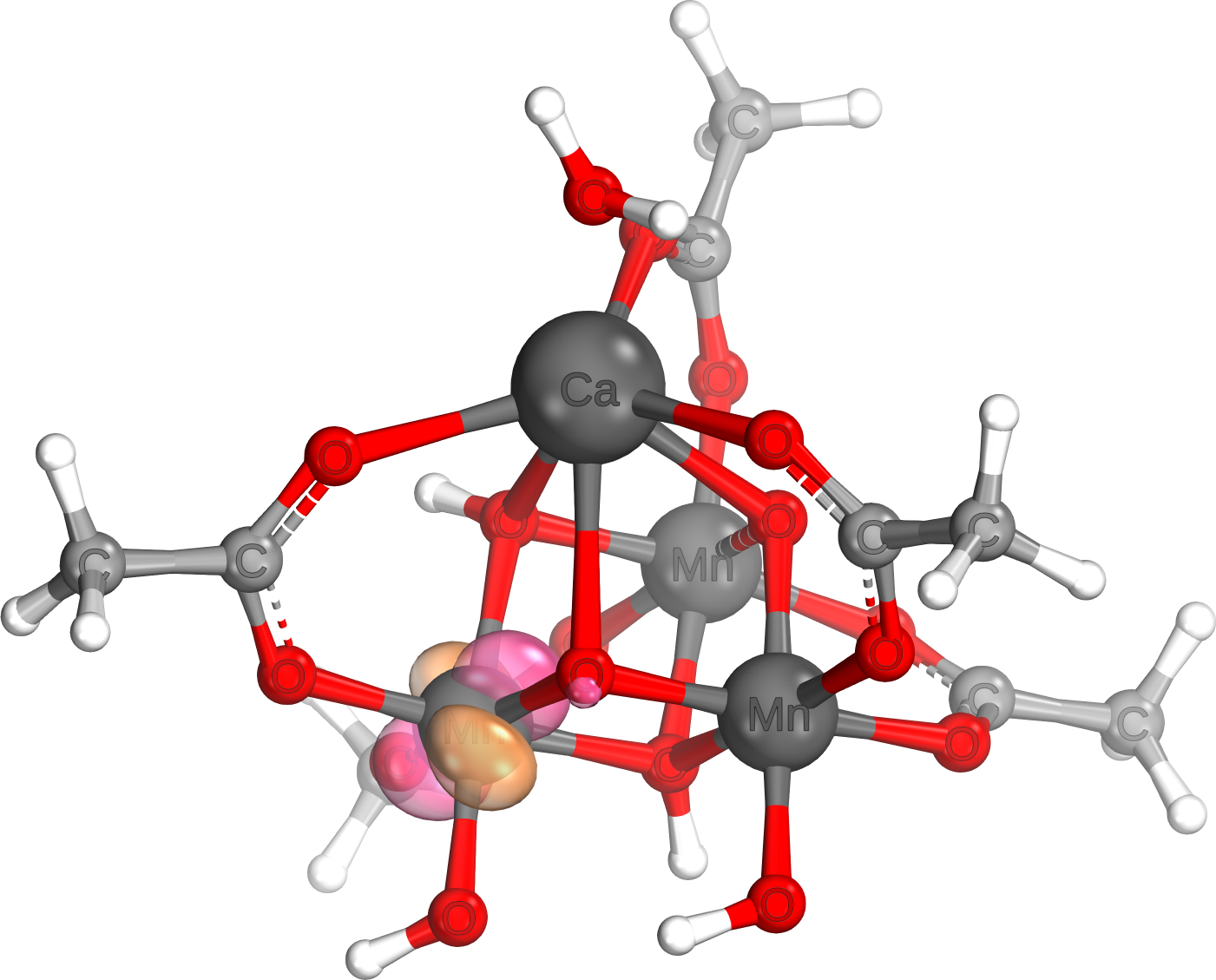}
    \includegraphics[width=0.245\textwidth]{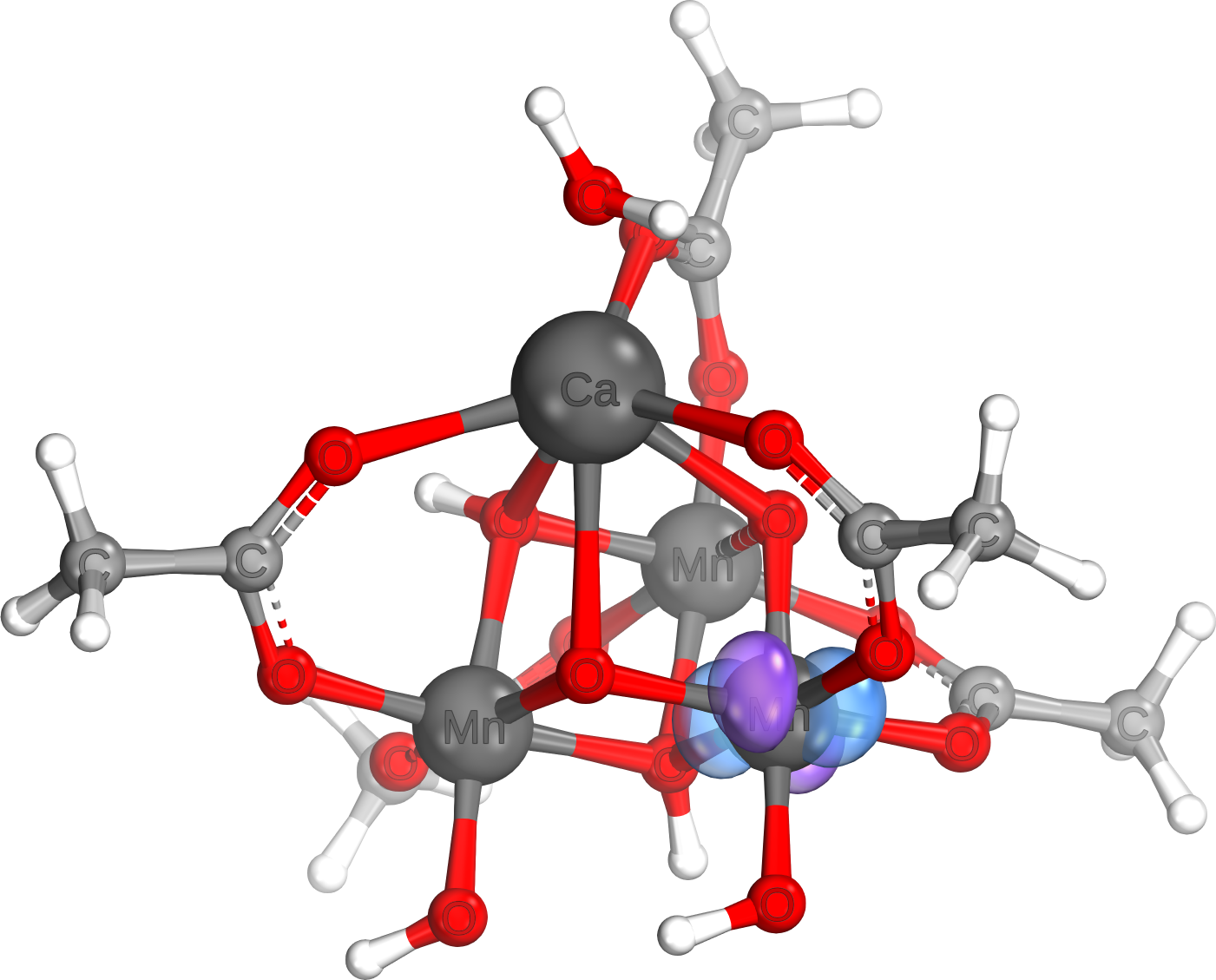} \\
    \includegraphics[width=0.245\textwidth]{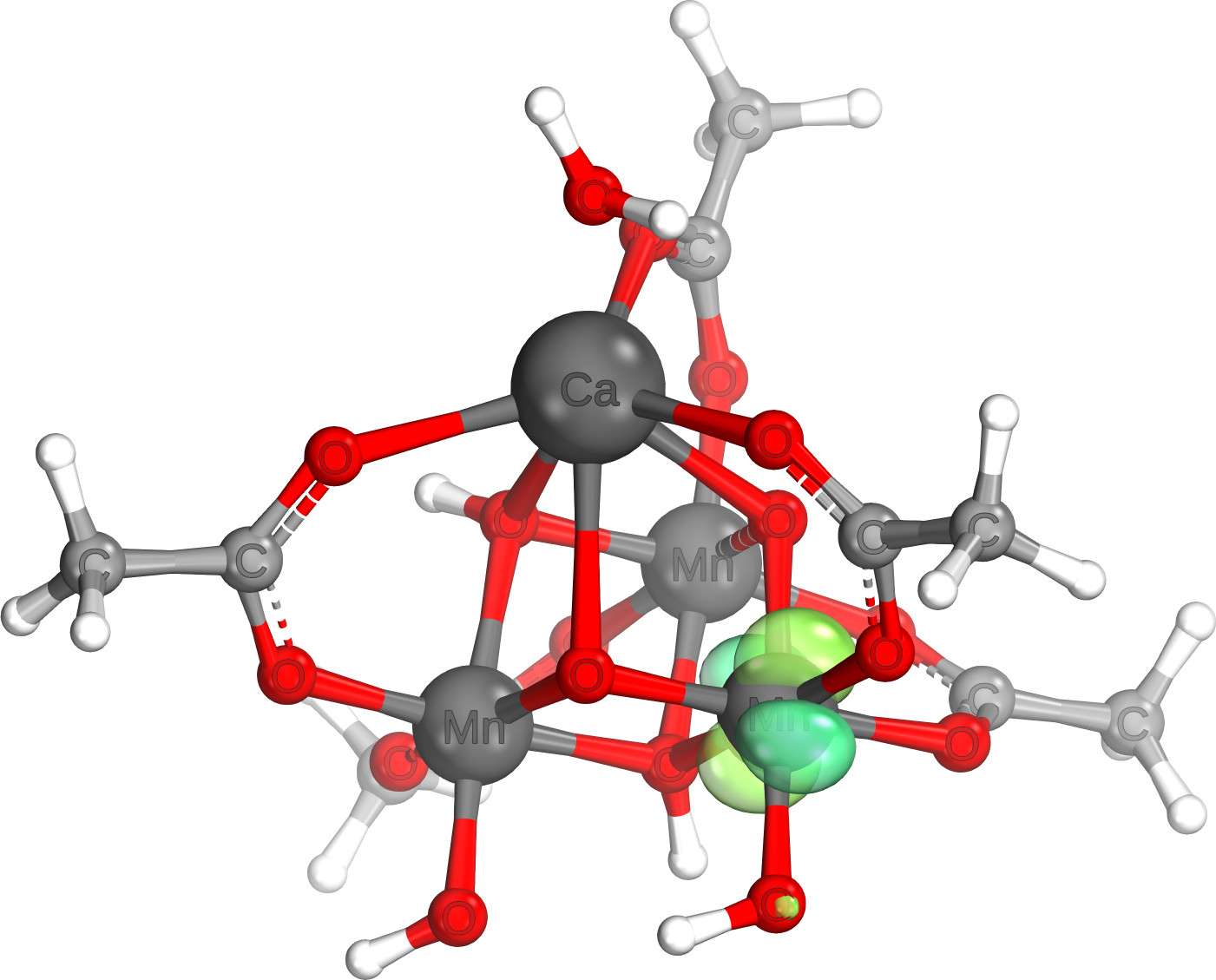}
    \includegraphics[width=0.245\textwidth]{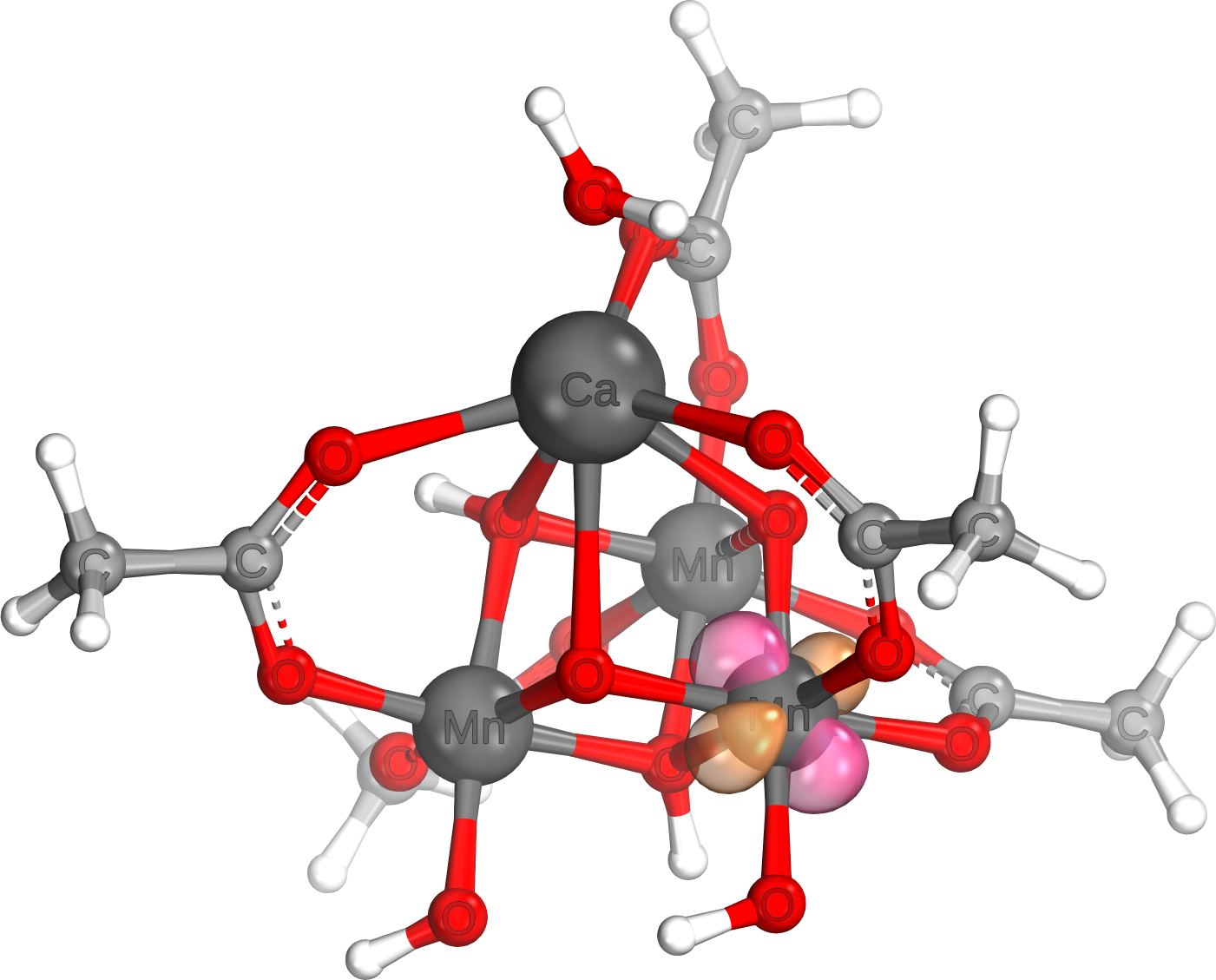}
    \includegraphics[width=0.245\textwidth]{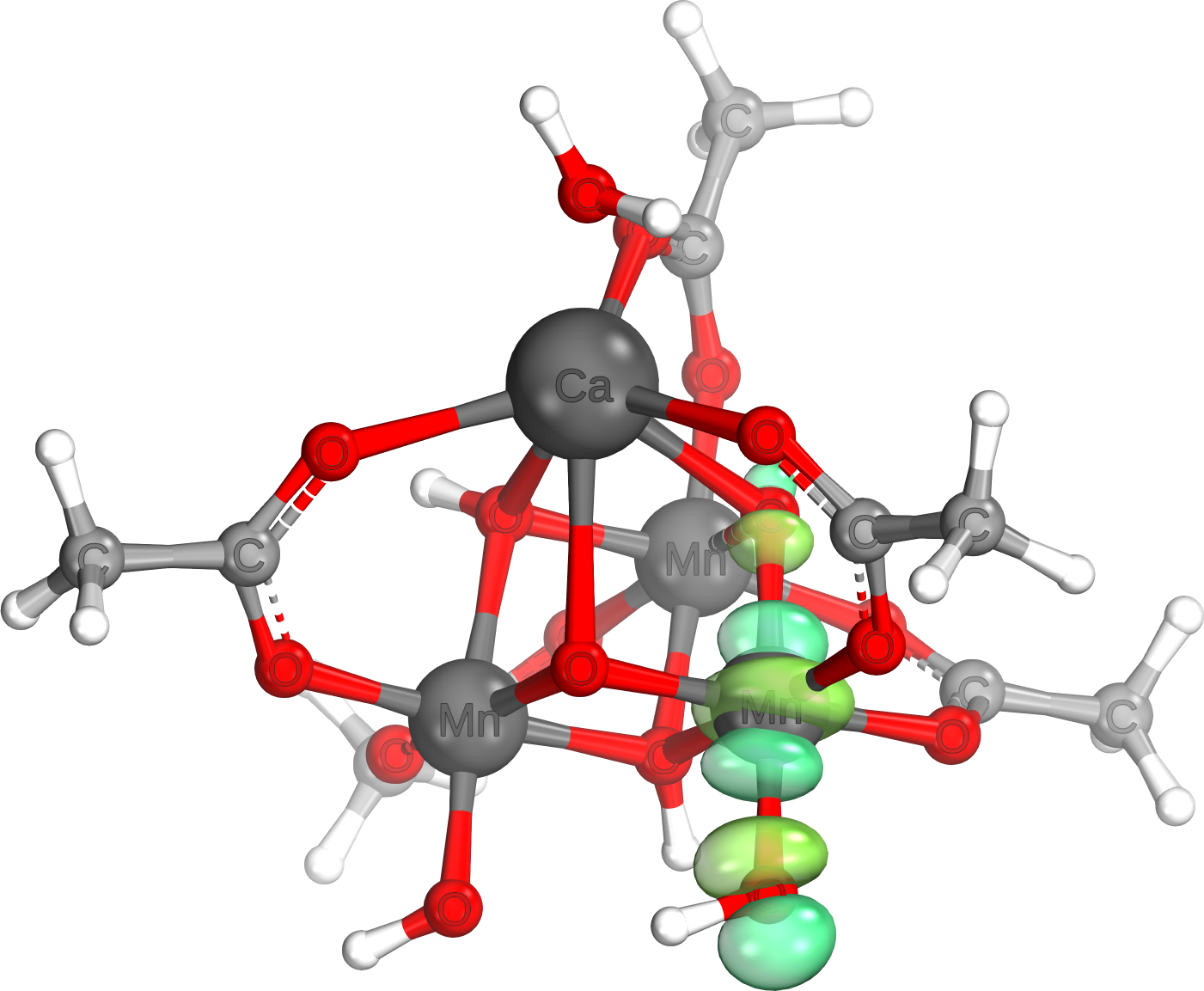}
    \includegraphics[width=0.245\textwidth]{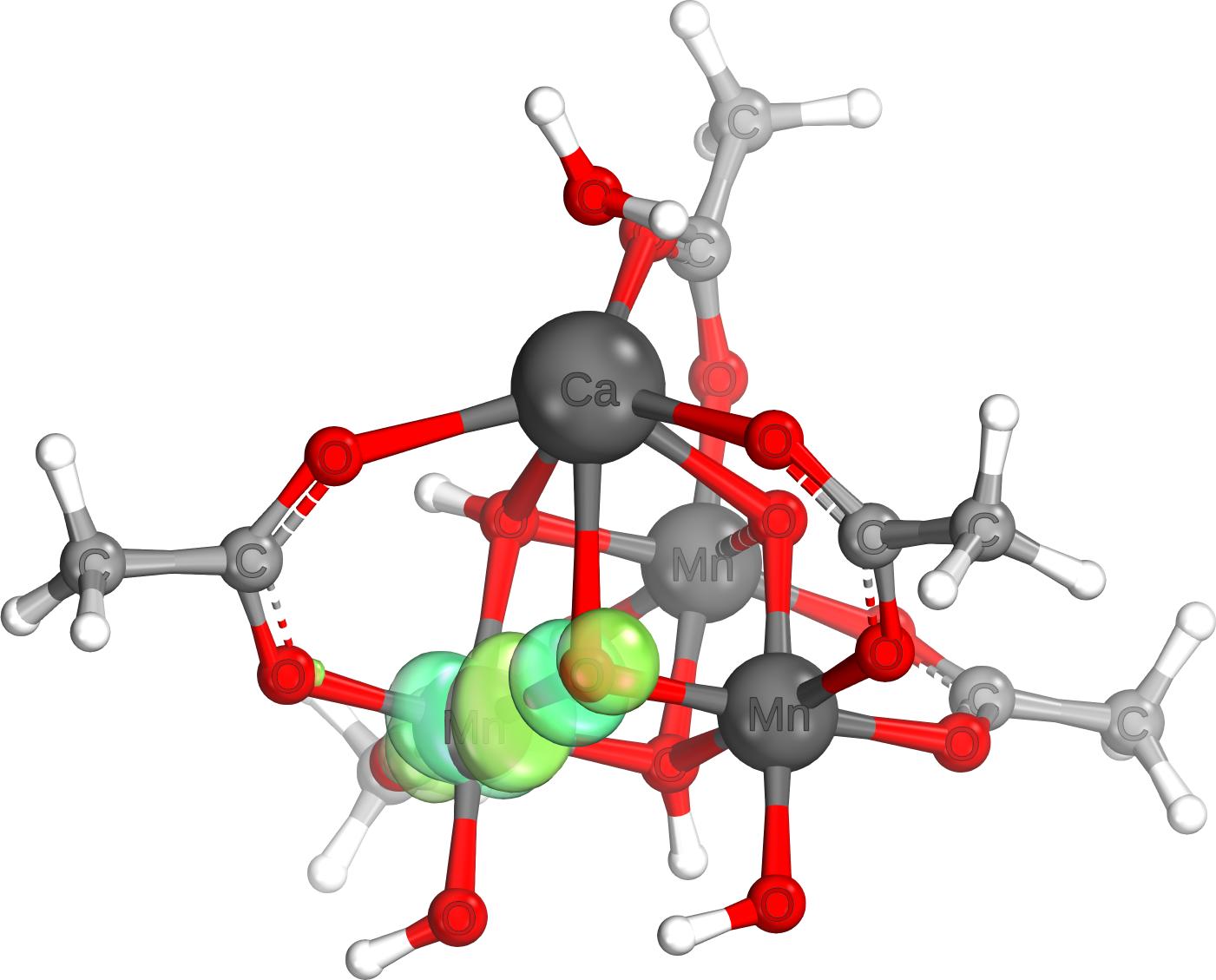} \\
    \includegraphics[width=0.245\textwidth]{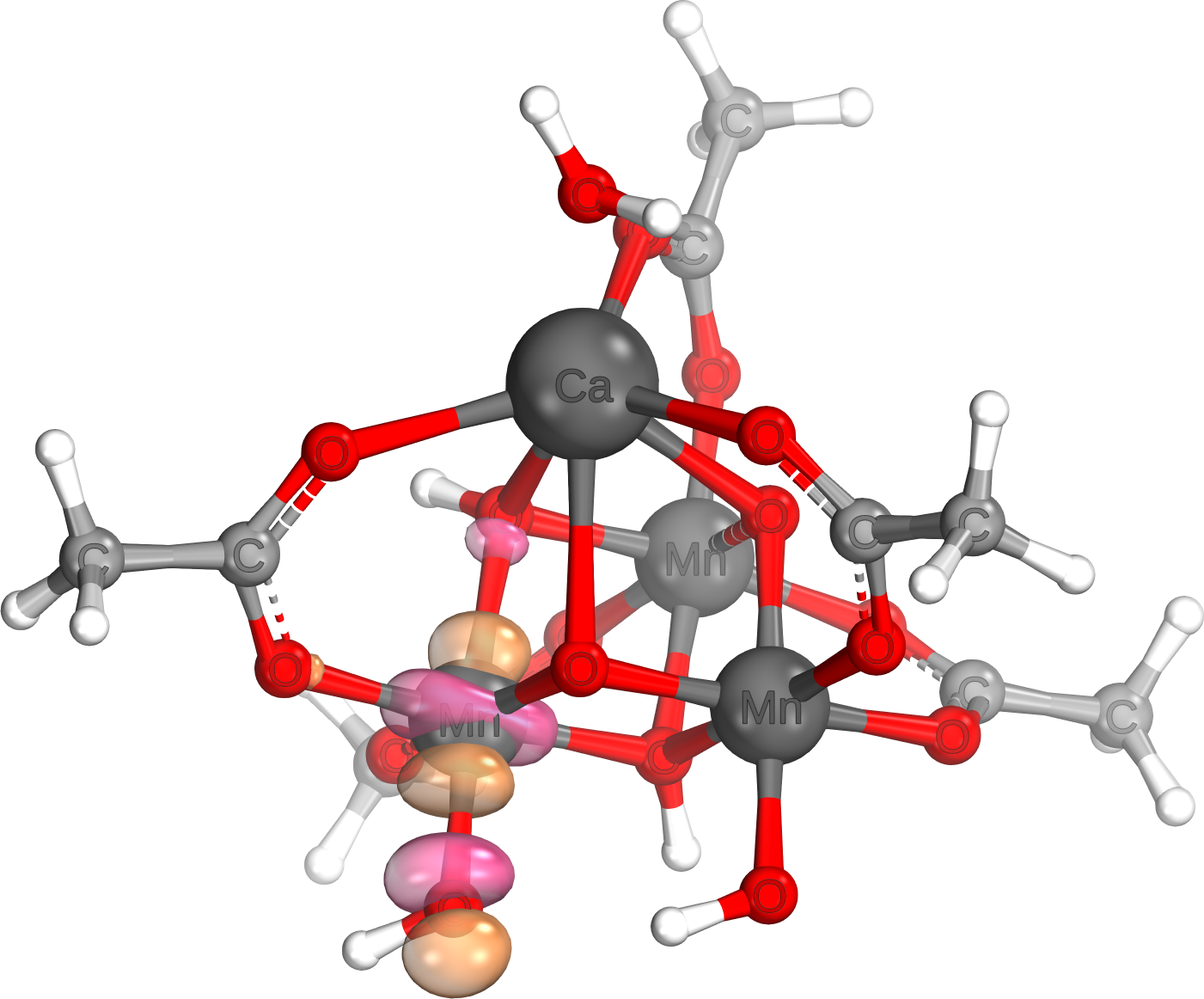}
    \includegraphics[width=0.245\textwidth]{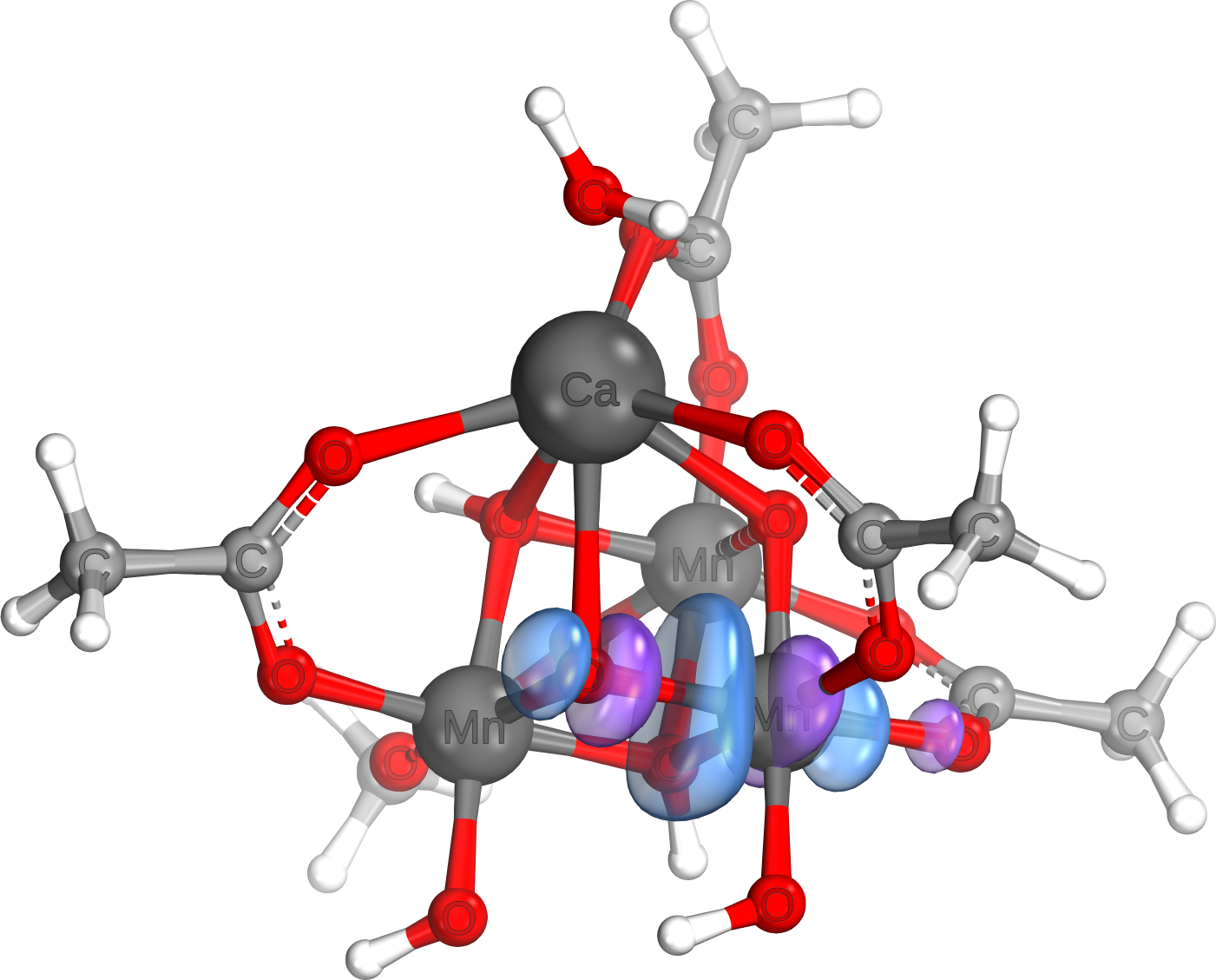}
    \includegraphics[width=0.245\textwidth]{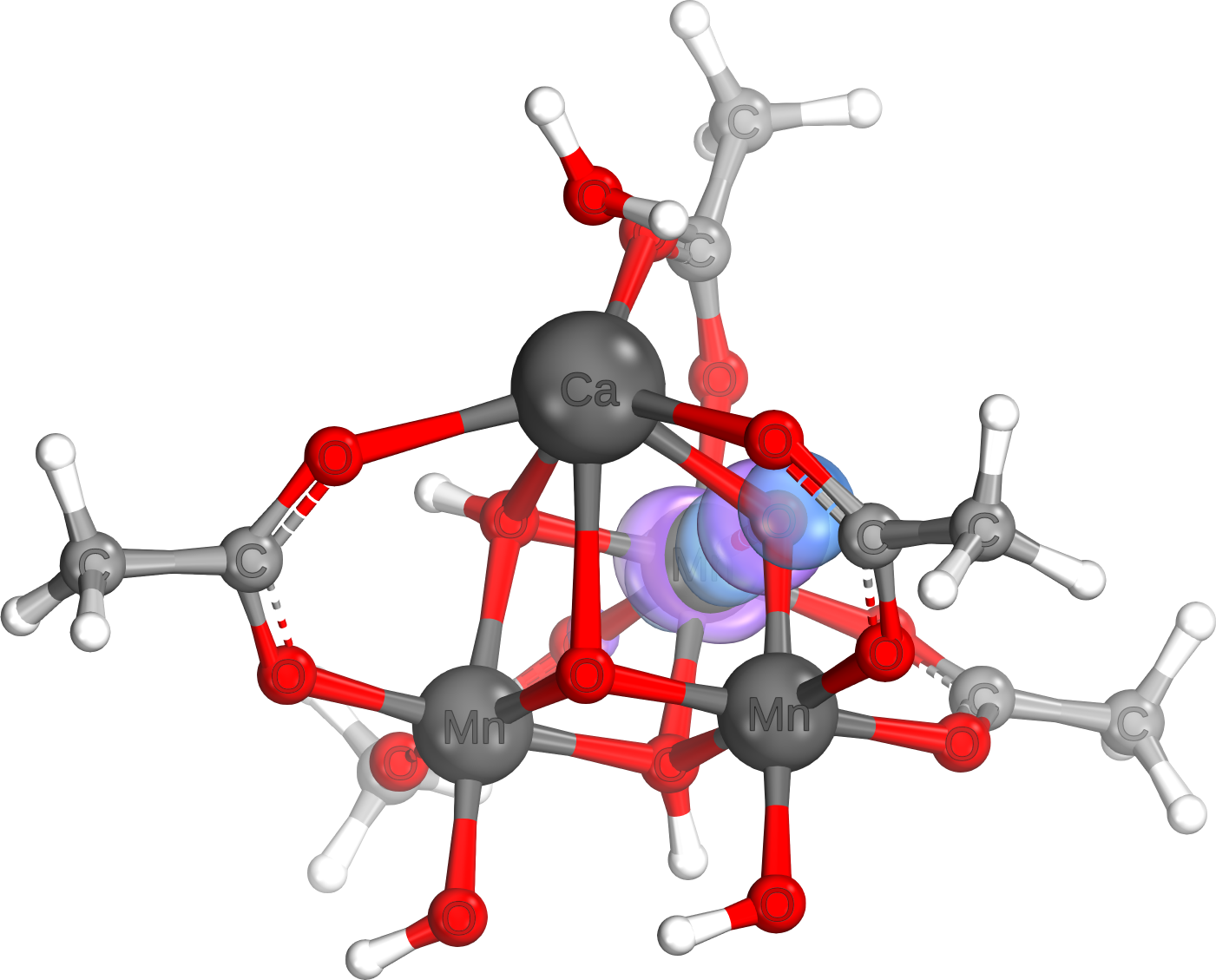}
    \includegraphics[width=0.245\textwidth]{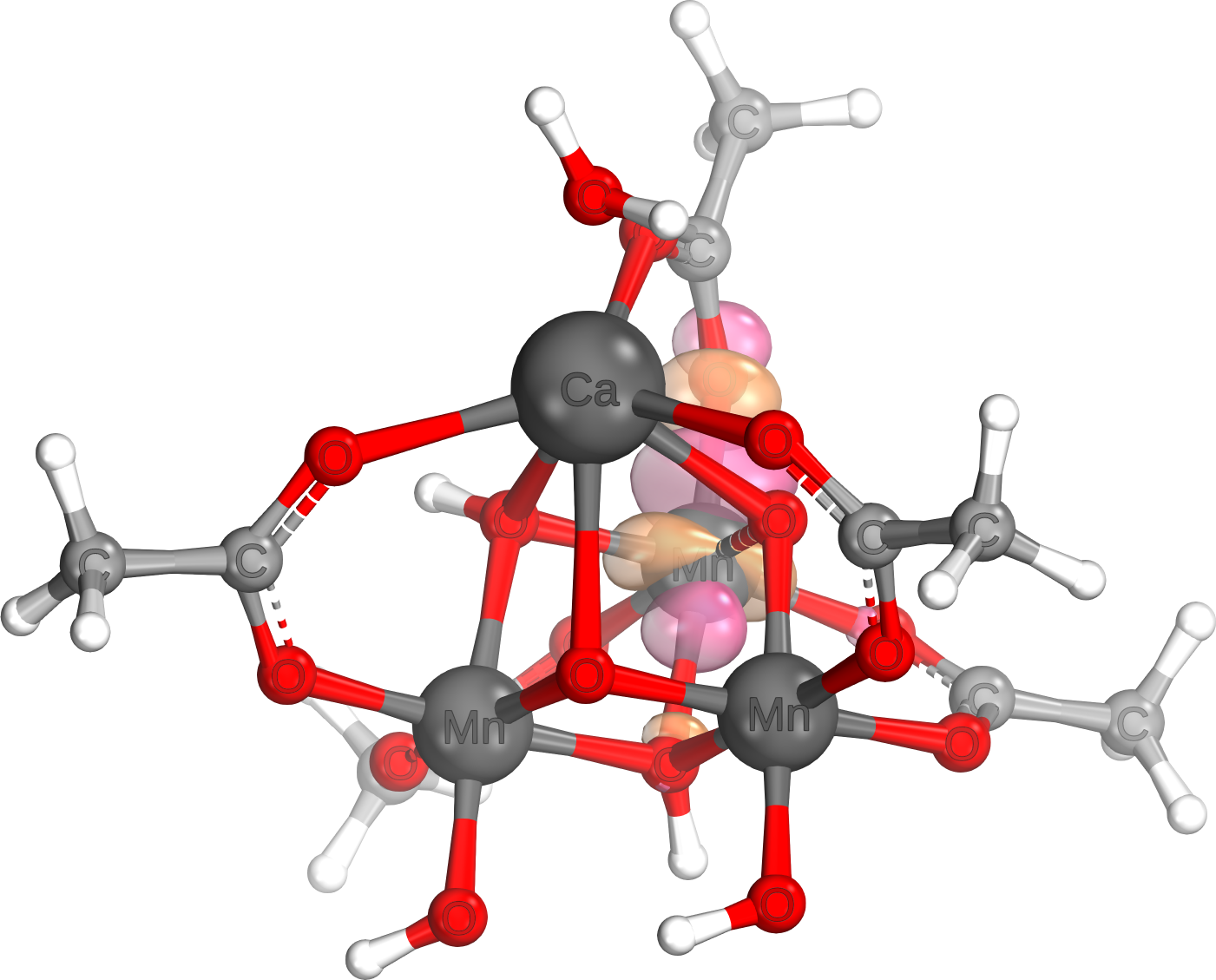} \\
    \caption{Split localised RASSCF(27, 2, 2; 9, 9, 6) orbitals used in the
    effective Hamiltonian mapping procedure.}
    \label{fig:split-local-rasorb}
\end{figure}

\section{Absolute Energies for the \ce{[CaMn3^{(IV)}O4]} Cubane}
\begin{table}[H]
    \caption{Total energies in Hartree obtained at various levels of theory for
    the protonated $\ce{[CaMn3^{(IV)}O4]}$ cubane discussed in the main text. CI roots with the
    same total spin are listed in descending order.}
    \resizebox{.99\hsize}{!}{
    { \footnotesize
    \begin{tabular}{@{}r ccccc@{}}
                      & $S_{\text{tot}} = \frac{1}{2}$  & $S_{\text{tot}} = \frac{3}{2}$ & $S_{\text{tot}} = \frac{5}{2}$ & $S_{\text{tot}} = \frac{7}{2}$ & $S_{\text{tot}} = \frac{9}{2}$ \\ \cmidrule{1-6}
        CASSCF(9, 9)  &                &                &                &                &                \\
                      & -5815.75979963 & -5815.76008349 & -5815.76025078 & -5815.76046090 & -5815.76070581 \\
                      & -5815.75963047 & -5815.75992670 & -5815.76005771 & -5815.76020897 &                \\
                      &                & -5815.75978378 & -5815.75982753 &                &                \\
                      &                & -5815.75955832 &                &                &                \\
        RASSCF(27, 24)&                &                &                &                &                \\
                      & -5816.01113429 & -5816.01141063 & -5816.01145458 & -5816.01151340 & -5816.01158419 \\
                      & -5816.01096566 & -5816.01115954 & -5816.01120145 & -5816.01125977 &                \\
                      &                & -5816.01098828 & -5816.01103209 &                &                \\
                      &                & -5816.01090401 &                &                &                \\
        +ftPBE        &                &                &                &                &                \\
                      & -5828.07716873 & -5828.07722610 & -5828.07700641 & -5828.07669678 & -5828.07628050 \\
                      & -5828.07705221 & -5828.07703292 & -5828.07681467 & -5828.07649991 &                \\
                      &                & -5828.07688971 & -5828.07669272 &                &                \\
                      &                & -5828.07687372 &                &                &                \\
        +ftBLYP       &                &                &                &                &                \\
                      & -5830.61378722 & -5830.61383019 & -5830.61358662 & -5830.61324374 & -5830.61278377 \\
                      & -5830.61367154 & -5830.61363695 & -5830.61339524 & -5830.61304674 &                \\
                      &                & -5830.61349280 & -5830.61327368 &                &                \\
                      &                & -5830.61348064 &                &                &                \\
        +ftSCAN\_E0   &                &                &                &                &                \\
                      & -5441.50160947 & -5441.50165847 & -5441.50105524 & -5441.50020835 & -5441.49908071 \\
                      & -5441.50135278 & -5441.50123951 & -5441.50064201 & -5441.49978106 &                \\
                      &                & -5441.50092332 & -5441.50037160 &                &                \\
                      &                & -5441.50089299 &                &                &                \\
    \end{tabular}
    }}
\end{table}
\nopagebreak

\section{Analytical Matrix for the 3J-3K Heisenberg Model}
\begin{table}[H]
  \vskip -2.5em
  \begin{equation}
    \hat{\mathcal{H}}_{3J3K} = \sum_{i < j}^{\{ A,B,C \}} J_{ij} (\hat{S}_i
    \cdot \hat{S}_j) - K_{ij} (\hat{S}_i \cdot \hat{S}_j)^2.
    \label{eq:heisenberg-general}
\end{equation}
\raggedright
 %\vskip -1.5em
Analytical matrix blocked by total spin with column labels
    $\ket{S_{\text{tot}}, S_{AB}}$. To enhance legibility, the substitutions
    $J_{3m} = J_{13} - J_{23}$, $J_{3p} = J_{13} + J_{23}$ and analogously for
    $K_{3m}$ and $K_{3p}$ were performed.
   {\footnotesize
\vskip 1em
    \begin{tabular}{@{}r cc@{}}
        $\hat{\mathcal{H}}_{3J3K}$  & $\ket{\frac{1}{2}, 1}$ & $\ket{\frac{1}{2}, 2}$ \\
        \cmidrule{1-3}
        $\bra{\frac{1}{2}, 1}$ & $\frac{1}{16} (-44 J_{12}-20 J_{3p}-121 K_{12} - 37 K_{3p})$ & $-\frac{1}{4} \sqrt{3} (2 J_{3m} + 7 K_{3m})                $ \\
        $\bra{\frac{1}{2}, 2}$ & $-\frac{1}{4} \sqrt{3} (2 J_{3m} + 7 K_{3m})               $ & $-\frac{3}{16} (4 J_{12} + 12 J_{3p} + 3 K_{12} + 31 K_{3p})$ \\
    \end{tabular}
% \end{table}

% \begin{table}[H]
%     \raggedright
% % \resizebox{.99\hsize}{!}{
\vskip 2em
    \begin{tabular}{@{}r cc@{}}
        $\hat{\mathcal{H}}_{3J3K}$  & $\ket{\frac{3}{2}, 0}$ & $\ket{\frac{3}{2}, 1}$ \\ %& $\ket{\frac{3}{2}, 2}$ & $\ket{\frac{3}{2}, 3}$ \\
        \cmidrule{1-3}
        $\bra{\frac{3}{2}, 0}$ & $-\frac{15}{16} (4 J_{12}+5 (3 K_{12} + K_{3p}))$ & $-\frac{5}{8} \sqrt{3} (2 J_{3m}+K_{3m})$                      \\ %& $-\frac{3 \sqrt{5} K_{3p}}{2}$                         &                                                                   \\
        $\bra{\frac{3}{2}, 1}$ & $-\frac{5}{8} \sqrt{3} (2 J_{3m}+K_{3m})$         & $\frac{1}{80} (-5 (44 J_{12}+8 J_{3p}+121 K_{12})-587 K_{3p})$ \\ %& $-2 \sqrt{\frac{3}{5}} (J_{3m}+2 K_{3m})$              & $-\frac{3 \sqrt{21} K_{3p}}{10}$                                  \\
        $\bra{\frac{3}{2}, 2}$ & $-\frac{3 \sqrt{5} K_{3p}}{2}$                    & $-2 \sqrt{\frac{3}{5}} (J_{3m}+2 K_{3m})$                      \\% & $-\frac{3}{16} (4 J_{12}+8 J_{3p}+3 K_{12}+29 K_{3p})$ & $-\frac{3}{8} \sqrt{\frac{7}{5}} (2 J_{3m}+9 K_{3m})$             \\
        $\bra{\frac{3}{2}, 3}$ & 0                                                 & $-\frac{3 \sqrt{21} K_{3p}}{10}$                               \\% & $-\frac{3}{8} \sqrt{\frac{7}{5}} (2 J_{3m}+9 K_{3m})$  & $\frac{9 J_{12}}{4}-3 J_{3p}-\frac{27}{80} (15 K_{12}+29 K_{3p})$ \\
        \\[-0.5em]
        $\hat{\mathcal{H}}_{3J3K}$  & $\ket{\frac{3}{2}, 2}$ & $\ket{\frac{3}{2}, 3}$ \\
        \cmidrule{1-3}
        $\bra{\frac{3}{2}, 0}$ & $-\frac{3 \sqrt{5} K_{3p}}{2}$                         & 0                                                                 \\
        $\bra{\frac{3}{2}, 1}$ & $-2 \sqrt{\frac{3}{5}} (J_{3m}+2 K_{3m})$              & $-\frac{3 \sqrt{21} K_{3p}}{10}$                                  \\
        $\bra{\frac{3}{2}, 2}$ & $-\frac{3}{16} (4 J_{12}+8 J_{3p}+3 K_{12}+29 K_{3p})$ & $-\frac{3}{8} \sqrt{\frac{7}{5}} (2 J_{3m}+9 K_{3m})$             \\
        $\bra{\frac{3}{2}, 3}$ & $-\frac{3}{8} \sqrt{\frac{7}{5}} (2 J_{3m}+9 K_{3m})$  & $\frac{9 J_{12}}{4}-3 J_{3p}-\frac{27}{80} (15 K_{12}+29 K_{3p})$ \\
    \end{tabular}
% % }
% \end{table}

% \begin{table}[H]
%     \raggedright
% %\resizebox{.99\hsize}{!}{
\vskip 2em
    \begin{tabular}{@{}r cc@{}}
        $\hat{\mathcal{H}}_{3J3K}$  & $\ket{\frac{5}{2}, 1}$ & $\ket{\frac{5}{2}, 2}$ \\ %& $\ket{\frac{5}{2}, 3}$ \\
        \cmidrule{1-3}
        $\bra{\frac{5}{2}, 1}$ & $\frac{1}{80} (-220 J_{12}+60 J_{3p}-605 K_{12}-297 K_{3p})$ & $\frac{3}{4} \sqrt{\frac{7}{5}} (K_{3m}-2 J_{3m})$      \\ %& $-\frac{3 \sqrt{14} K_{3p}}{5}$                              \\
        $\bra{\frac{5}{2}, 2}$ & $\frac{3}{4} \sqrt{\frac{7}{5}} (K_{3m}-2 J_{3m})$           & $\frac{1}{16} (-12 J_{12}-4 J_{3p}-9 K_{12}-77 K_{3p})$ \\ %& $-2 \sqrt{\frac{2}{5}} (J_{3m}+2 K_{3m})$                    \\
        $\bra{\frac{5}{2}, 3}$ & $-\frac{3 \sqrt{14} K_{3p}}{5}$                              & $-2 \sqrt{\frac{2}{5}} (J_{3m}+2 K_{3m})$               \\ %& $\frac{1}{80} (180 J_{12}-140 J_{3p}-405 K_{12}-373 K_{3p})$ \\
        \\[-0.5em]
        $\hat{\mathcal{H}}_{3J3K}$   & $\ket{\frac{5}{2}, 3}$ \\
        \cmidrule{1-2}
        $\bra{\frac{5}{2}, 1}$  & $-\frac{3 \sqrt{14} K_{3p}}{5}$                              \\
        $\bra{\frac{5}{2}, 2}$  & $-2 \sqrt{\frac{2}{5}} (J_{3m}+2 K_{3m})$                    \\
        $\bra{\frac{5}{2}, 3}$  & $\frac{1}{80} (180 J_{12}-140 J_{3p}-405 K_{12}-373 K_{3p})$ \\
    \end{tabular}
% %}
% \end{table}

% \begin{table}[H]
%     \raggedright
\vskip 2em
    \begin{tabular}{@{}r cc@{}}
        $\hat{\mathcal{H}}_{3J3K}$  & $\ket{\frac{7}{2}, 2}$ & $\ket{\frac{7}{2}, 3}$ \\
        \cmidrule{1-3}
        $\bra{\frac{7}{2}, 2}$ & $-\frac{3}{16} (4 J_{12}-8 J_{3p}+3 (K_{12}+7 K_{3p}))$ & $\frac{3}{8} \sqrt{3} (3 K_{3m}-2 J_{3m})$    \\
        $\bra{\frac{7}{2}, 3}$ & $\frac{3}{8} \sqrt{3} (3 K_{3m}-2 J_{3m})$              & $\frac{9}{16} (4 J_{12}-3 (3 K_{12}+K_{3p}))$ \\
    \end{tabular}
% \end{table}

% \begin{table}[H]
%     \raggedright
\vskip 2em
    \begin{tabular}{@{}r c@{}}
        $\hat{\mathcal{H}}_{3J3K}$  & $\ket{\frac{9}{2}, 3}$ \\
        \cmidrule{1-2}
        $\bra{\frac{9}{2}, 3}$ & $\frac{9}{16} (4 J_{12}+4 J_{3p}-9 (K_{12}+K_{3p}))$ \\
    \end{tabular}
  }
\end{table}

\pagebreak
\section{Effective Hamiltonian from CASSCF(9,9)}
Left column: effective Hamiltonian obtained from CASSCF(9,9), right column:
model Hamiltonian (\ref{eq:heisenberg-general}) with CASSCF(9,9) parameters
from Table 4 plugged in.
Both blocked by total spin with column labels $\ket{S_{\text{tot}}, S_{AB}}$.
All values are given in units of $\SI{}{\per\cm}$ relative to
$\ket{\frac{9}{2}, 3}$.

\begin{minipage}[t]{0.48\textwidth}
\raggedright
\begin{table}[H]
    \raggedright
    \begin{tabular}{@{}r cc@{}}
        $\hat{\mathcal{H}}_\mathrm{eff}^\mathrm{CASSCF}$  & $\ket{\frac{1}{2}, 1}$ & $\ket{\frac{1}{2}, 2}$ \\
        \cmidrule{1-3}
        $\bra{\frac{1}{2}, 1}$ & 200 &   7 \\[2pt]
        $\bra{\frac{1}{2}, 2}$ &   7 & 235 \\
    \end{tabular}
\end{table}

\begin{table}[H]
    \raggedright
    \begin{tabular}{@{}r cccc@{}}
        $\hat{\mathcal{H}}_\mathrm{eff}^\mathrm{CASSCF}$  & $\ket{\frac{3}{2}, 0}$ & $\ket{\frac{3}{2}, 1}$ & $\ket{\frac{3}{2}, 2}$ & $\ket{\frac{3}{2}, 3}$ \\[2pt]
        \cmidrule{1-5}
        $\bra{\frac{3}{2}, 0}$ & 147 &  17 &     &     \\[2pt]
        $\bra{\frac{3}{2}, 1}$ &  17 & 165 &  12 &     \\[2pt]
        $\bra{\frac{3}{2}, 2}$ &     &  12 & 199 &   7 \\[2pt]
        $\bra{\frac{3}{2}, 3}$ &     &     &   7 & 251 \\
    \end{tabular}
\end{table}

\begin{table}[H]
    \raggedright
    \begin{tabular}{@{}r ccc@{}}
        $\hat{\mathcal{H}}_\mathrm{eff}^\mathrm{CASSCF}$  & $\ket{\frac{5}{2}, 1}$ & $\ket{\frac{5}{2}, 2}$ & $\ket{\frac{5}{2}, 3}$ \\
        \cmidrule{1-4}
        $\bra{\frac{5}{2}, 1}$ & 105 &   14 &     \\[2pt]
        $\bra{\frac{5}{2}, 2}$ &  14 &  139 &  10 \\[2pt]
        $\bra{\frac{5}{2}, 3}$ &     &   10 & 191 \\
    \end{tabular}
\end{table}

\begin{table}[H]
    \raggedright
    \begin{tabular}{@{}r cc@{}}
        $\hat{\mathcal{H}}_\mathrm{eff}^\mathrm{CASSCF}$  & $\ket{\frac{7}{2}, 2}$ & $\ket{\frac{7}{2}, 3}$ \\
        \cmidrule{1-3}
        $\bra{\frac{7}{2}, 2}$ &  56 &  10 \\[2pt]
        $\bra{\frac{7}{2}, 3}$ &  10 & 107 \\
    \end{tabular}
\end{table}
\end{minipage}
\hfill
\begin{minipage}[t]{0.48\textwidth}
  \begin{table}[H]
    \raggedright
    \begin{tabular}{@{}r cc@{}}
        $\hat{\mathcal{H}}_{3J3K}^\mathrm{CASSCF}$  & $\ket{\frac{1}{2}, 1}$ & $\ket{\frac{1}{2}, 2}$ \\
        \cmidrule{1-3}
        $\bra{\frac{1}{2}, 1}$ & 200 &   7 \\[2pt]
        $\bra{\frac{1}{2}, 2}$ &   7 & 235 \\
    \end{tabular}
\end{table}

\begin{table}[H]
    \raggedright
    \begin{tabular}{@{}r cccc@{}}
        $\hat{\mathcal{H}}_{3J3K}^\mathrm{CASSCF}$  & $\ket{\frac{3}{2}, 0}$ & $\ket{\frac{3}{2}, 1}$ & $\ket{\frac{3}{2}, 2}$ & $\ket{\frac{3}{2}, 3}$ \\
        \cmidrule{1-5}
        $\bra{\frac{3}{2}, 0}$ & 147 &  17 &     &     \\[2pt]
        $\bra{\frac{3}{2}, 1}$ &  17 & 165 &  12 &     \\[2pt]
        $\bra{\frac{3}{2}, 2}$ &     &  12 & 199 &   7 \\[2pt]
        $\bra{\frac{3}{2}, 3}$ &     &     &   7 & 251 \\
    \end{tabular}
\end{table}

\begin{table}[H]
    \raggedright
    \begin{tabular}{@{}r ccc@{}}
        $\hat{\mathcal{H}}_{3J3K}^\mathrm{CASSCF}$  & $\ket{\frac{5}{2}, 1}$ & $\ket{\frac{5}{2}, 2}$ & $\ket{\frac{5}{2}, 3}$ \\
        \cmidrule{1-4}
        $\bra{\frac{5}{2}, 1}$ & 105 &  14 &     \\[2pt]
        $\bra{\frac{5}{2}, 2}$ &  14 & 139 &  10 \\[2pt]
        $\bra{\frac{5}{2}, 3}$ &     &  10 & 191 \\
    \end{tabular}
\end{table}

\begin{table}[H]
    \raggedright
    \begin{tabular}{@{}r cc@{}}
        $\hat{\mathcal{H}}_{3J3K}^\mathrm{CASSCF}$  & $\ket{\frac{7}{2}, 2}$ & $\ket{\frac{7}{2}, 3}$ \\
        \cmidrule{1-3}
        $\bra{\frac{7}{2}, 2}$ & 56 &  10 \\[2pt]
        $\bra{\frac{7}{2}, 3}$ & 10 & 107 \\
    \end{tabular}
\end{table}
\end{minipage}

\pagebreak
\section{Effective Hamiltonian from \mbox{RASSCF(27, 2, 2; 9, 9, 6)}}
Left column: effective Hamiltonian obtained from RASSCF(27, 2, 2; 9, 9, 6),
right column: model Hamiltonian (\ref{eq:heisenberg-general}) with
RASSCF(27, 2, 2; 9, 9, 6) parameters from Table 4 plugged in.
Both blocked by total spin with column labels $\ket{S_{\text{tot}}, S_{AB}}$.
All values are given in units of $\SI{}{\per\cm}$ relative to
$\ket{\frac{9}{2}, 3}$.

\begin{minipage}[t]{0.48\textwidth}
\raggedright
\begin{table}[H]
    \raggedright
    \begin{tabular}{@{}r cc@{}}
        $\hat{\mathcal{H}}_\mathrm{eff}^\mathrm{RASSCF}$  & $\ket{\frac{1}{2}, 1}$ & $\ket{\frac{1}{2}, 2}$ \\
        \cmidrule{1-3}
        $\bra{\frac{1}{2}, 1}$ & 108 &  16 \\[2pt]
        $\bra{\frac{1}{2}, 2}$ &  16 & 127 \\
    \end{tabular}
\end{table}

\begin{table}[H]
    \raggedright
    \begin{tabular}{@{}r cccc@{}}
        $\hat{\mathcal{H}}_\mathrm{eff}^\mathrm{RASSCF}$  & $\ket{\frac{3}{2}, 0}$ & $\ket{\frac{3}{2}, 1}$ & $\ket{\frac{3}{2}, 2}$ & $\ket{\frac{3}{2}, 3}$ \\
        \cmidrule{1-5}
        $\bra{\frac{3}{2}, 0}$ & 79 & 40 &     &     \\[2pt]
        $\bra{\frac{3}{2}, 1}$ & 40 & 89 &  28 &     \\[2pt]
        $\bra{\frac{3}{2}, 2}$ &    & 28 & 107 &  16 \\[2pt]
        $\bra{\frac{3}{2}, 3}$ &    &    &  16 & 136 \\
    \end{tabular}
\end{table}

\begin{table}[H]
    \raggedright
    \begin{tabular}{@{}r ccc@{}}
        $\hat{\mathcal{H}}_\mathrm{eff}^\mathrm{RASSCF}$  & $\ket{\frac{5}{2}, 1}$ & $\ket{\frac{5}{2}, 2}$ & $\ket{\frac{5}{2}, 3}$ \\
        \cmidrule{1-4}
        $\bra{\frac{5}{2}, 1}$ &  56 &  33 &     \\[2pt]
        $\bra{\frac{5}{2}, 2}$ &  33 &  75 &  23 \\[2pt]
        $\bra{\frac{5}{2}, 3}$ &     &  23 & 103 \\
    \end{tabular}
\end{table}

\begin{table}[H]
    \raggedright
    \begin{tabular}{@{}r cc@{}}
        $\hat{\mathcal{H}}_\mathrm{eff}^\mathrm{RASSCF}$  & $\ket{\frac{7}{2}, 2}$ & $\ket{\frac{7}{2}, 3}$ \\
        \cmidrule{1-3}
        $\bra{\frac{7}{2}, 2}$ & 30 & 24 \\[2pt]
        $\bra{\frac{7}{2}, 3}$ & 24 & 57 \\
    \end{tabular}
\end{table}
\end{minipage}
\hfill
\begin{minipage}[t]{0.48\textwidth}
\raggedright
\begin{table}[H]
    \raggedright
    \begin{tabular}{@{}r cc@{}}
        $\hat{\mathcal{H}}_{3J3K}^\mathrm{RASSCF}$  & $\ket{\frac{1}{2}, 1}$ & $\ket{\frac{1}{2}, 2}$ \\
        \cmidrule{1-3}
        $\bra{\frac{1}{2}, 1}$ & 108 &  16 \\[2pt]
        $\bra{\frac{1}{2}, 2}$ &  16 & 127 \\
    \end{tabular}
\end{table}

\begin{table}[H]
    \raggedright
    \begin{tabular}{@{}r cccc@{}}
        $\hat{\mathcal{H}}_{3J3K}^\mathrm{RASSCF}$  & $\ket{\frac{3}{2}, 0}$ & $\ket{\frac{3}{2}, 1}$ & $\ket{\frac{3}{2}, 2}$ & $\ket{\frac{3}{2}, 3}$ \\
        \cmidrule{1-5}
        $\bra{\frac{3}{2}, 0}$ &  79  &  40  &   1 &     \\[2pt]
        $\bra{\frac{3}{2}, 1}$ &  40  &  89  &  28 &     \\[2pt]
        $\bra{\frac{3}{2}, 2}$ &   1  &  28  & 108 &  16 \\[2pt]
        $\bra{\frac{3}{2}, 3}$ &      &      &  16 & 137 \\
    \end{tabular}
\end{table}

\begin{table}[H]
    \raggedright
    \begin{tabular}{@{}r ccc@{}}
        $\hat{\mathcal{H}}_{3J3K}^\mathrm{RASSCF}$  & $\ket{\frac{5}{2}, 1}$ & $\ket{\frac{5}{2}, 2}$ & $\ket{\frac{5}{2}, 3}$ \\
        \cmidrule{1-4}
        $\bra{\frac{5}{2}, 1}$ &  56  &  33  &     \\[2pt]
        $\bra{\frac{5}{2}, 2}$ &  33  &  75  &  23 \\[2pt]
        $\bra{\frac{5}{2}, 3}$ &      &  23  & 103 \\
    \end{tabular}
\end{table}

\begin{table}[H]
    \raggedright
    \begin{tabular}{@{}r cc@{}}
        $\hat{\mathcal{H}}_{3J3K}^\mathrm{RASSCF}$  & $\ket{\frac{7}{2}, 2}$ & $\ket{\frac{7}{2}, 3}$ \\
        \cmidrule{1-3}
        $\bra{\frac{7}{2}, 2}$ &  30  &  24 \\[2pt]
        $\bra{\frac{7}{2}, 3}$ &  24  &  58 \\
    \end{tabular}
\end{table}
\end{minipage}

\pagebreak
\section{Effective Hamiltonians with Localised Active Space State Interaction}
In the localised active space (LAS) framework, the wave function is expressed
    as an anti-symmetrised tensor product of fragment wave functions $\ket{n_i}$:
\begin{equation}
    \ket{\mathrm{LAS}} = \bigwedge_{i}^{\mathrm{fragments}} \ket{n_i},
    \label{si:las-ansatz}
\end{equation}
    where it was assumed that the doubly occupied orbitals are frozen.
The $\ket{n_i}$ are fragment wave functions with well-defined subsystem
    quantum numbers.
Missing interspace correlation is obtained by forming linear combinations
    of LAS-states in LAS state interaction (LASSI).
Recently, a semi-automatic procedure\cite{agarawal2024} for LASSI was
    introduced that operates with the number of inter-fragment excitations $r$ and
    intra-fragment excited states $q$, denoted LASSI[r, q].
In the limit $r,q \rightarrow \infty$ the method is equivalent to a CAS ansatz.

For active spaces comprising multiple magnetic sites, equation
    \eqref{si:las-ansatz} closely resembles the uncoupled basis of a spin
    Hamiltonian $\ket{j_1 m_1; \text{ }j_2 m_2; \text{ }\ldots; \text{ }j_i m_i}$
    where the conserved subsystem quantum numbers on site $i$ are $j_i$ and $m_i$.
Since $m_\mathrm{total} = \sum_i m_i$ is a good quantum number for
    the models we consider, the Hamiltonian in the uncoupled basis is blocked
    by $m_\mathrm{total}$.
Choosing the sub-block with the smallest $m_\mathrm{total}$ provides the
    smallest non-redundant form of the Hamiltonian that can be used to obtain the
    coupling parameters.
If each $\ket{n_i}$ corresponds to a magnetic site, LASSI[0, 1] recovers up to
    a phase the Clebsch-Gordan coefficients occurring in the tensor product space
    reduction of $\ket{j_1 m_1; \text{ }j_2 m_2; \text{ }\ldots; \text{ }j_i m_i}$.
Consequently, rather than using the recoupling procedure presented in the main
    manuscript, general LASSI[r, q] states can be projected onto the
    LASSI[0, 1] manifold to form the effective Hamiltonian.

\bibliography{../references}